\documentclass[usenatbib]{mnras}
\usepackage{mathptmx}
\usepackage{color}
\usepackage{graphics,graphicx,amssymb,amsmath,fixltx2e,natbib}
\usepackage{epsfig}

\def\gtrsim{\lower 2pt \hbox{$\, \buildrel {\scriptstyle >}\over
{\scriptstyle \sim}\,$}}
\def\lesssim{\lower 2pt \hbox{$\, \buildrel {\scriptstyle <}\over
{\scriptstyle \sim}\,$}}

\def\xmm{{\sl XMM-Newton}}

\def\suzaku{{\sl Suzaku}}
\def\chandra{{\sl Chandra}}

\def\suzaku{{\sl Suzaku}}

\def\wise{{\sl WISE}}
\def\meerkat{{\sl MeerKAT}}

\def\suzaku{{\sl Suzaku}}
\def\micron{{~$\mu$m}}

\def\xs{G0.17-0.41}
\def\rxs{G359.55+0.16}

\begin{document}

\title{Chandra large-scale mapping of the Galactic center: Probing high-energy structures around the central molecular zone}
\author[Q. Daniel Wang et al.]{Q. Daniel Wang\\ 
$^{1}$Department of Astronomy, University of Massachusetts,  Amherst, MA 01003, USA (wqd@umass.edu)\\
}
\maketitle
\label{firstpage}
\begin{abstract}

Recent observations have revealed interstellar features that apparently connect energetic activity in the central region of our Galaxy to its halo. The nature of these features, however, remains largely uncertain. We present a \chandra\ mapping of the central $2^\circ \times 4^\circ$ field of the Galaxy,  revealing a complex of X-ray-emitting threads plus plume-like structures emerging from the Galactic center (GC). This mapping shows that the northern plume or fountain is offset from a well-known radio lobe (or the GCL), which however may represent a foreground HII region, and that the southern plume is well wrapped by a corresponding radio lobe recently discovered by {\sl MeerKAT}. In particular, we find that a distinct X-ray thread, G0.17-0.41, is embedded well within a nonthermal radio filament, which is locally inflated. This thread with a width of $\sim 1\farcs6$ (FWHM) is  $\sim  2\farcm6$ or 6~pc long at the distance of the GC and has a spectrum that can be characterized by a power law or an optically-thin thermal plasma with temperature $\gtrsim 3$~keV. The X-ray-emitting material is likely confined within a  strand of magnetic field with its strength $\gtrsim 1$~mG, not unusual in such radio filaments. These morphological and spectral properties of the radio/X-ray association suggest that magnetic field re-connection is the energy source. Such re-connection events are probably common when flux tubes of antiparallel magnetic fields collide and/or become twisted in and around the diffuse X-ray plumes, representing blowout superbubbles driven by young massive stellar clusters  in the GC. The understanding of the process, theoretically predicted in analog  to solar flares, can have strong implications for the study of interstellar hot plasma heating, cosmic-ray acceleration and turbulence.
\end{abstract}

\begin{keywords}
{Galaxy: center, evolution, ISM: magnetic fields, bubbles, jets and outflows, X-rays: general}
\end{keywords}

\section{Introduction}\label{s:int}

Because of its proximity ($D \approx 8$~kpc; 1$^{\prime\prime}$=0.039 pc), the Galactic center (GC)  of our Galaxy provides a critical test bed for developing our understanding of astrophysical processes under similar conditions throughout the cosmic time \citep[e.g.,][]{Morris1996,Kruijssen2013,Bland-Hawthorn2016,Armillotta2020}. The environment of the GC is characterized by the high temperature, density, turbulent velocity, and magnetic field of the interstellar medium (ISM), in addition to the strong gravitational tidal force around Sgr A* -- the central massive black hole of our Galaxy. 
Although Sgr A* is quiescent at present \citep[e.g.,][]{Wang2013}, it was evidently active in recent past, which has greatly influenced the structure, kinematics, and/or ionization of the ISM of the Galaxy and beyond \cite[e.g.,][]{Su2010,Ponti2015,Bland-Hawthorn2016,Koyama2018,DiTeodoro2018,Ashley2020}. 
Around Sgr A* are some of the most massive young stellar clusters in the Galaxy: Arches and Quintuplet clusters, as well as the GC cluster, in the age range of 2-7~Myr \citep[e.g.,][]{Figer2002}.
Dense gas is found predominately in the so-called central molecular zone (CMZ), including a twist ring or streams of giant molecular clouds (GMCs) orbiting around Sgr A* \citep{Kruijssen2015,Ginsburg2016,Barnes2017,Libralato2020}. 
These clouds have a mean temperature of $\sim 60-120$~K and is extremely turbulent (e.g., FWHM $\sim 20-50 {\rm~km~s^{-1}}$ on scales of a few parsecs; \citealt{Ginsburg2016}). Bursts of star formation are ongoing, but only in the most massive clouds such as Sgr B2
\citep[e.g., ][]{Barnes2017}.  

Recently, an apparent connection of such activity in the GC to large-scale energetic structures in the Galactic bulge or halo has been revealed in radio and X-ray. \suzaku\  and \xmm\ observations have shown hot plasma plumes (or fountains), emanating  from the center to the high Galactic latitudes \citep{Nakashima2013,Nakashima2019,Ponti2019}, while  in the same general regions the \meerkat\ mapping has revealed exquisite details of
radio-emitting features  \citep{Heywood2019}. Most noticeable are large-scale diffuse radio lobes, as well as numerous narrow filaments or their bundles, predominately ``radiating'' away from the most active portion of the CMZ. They are seen not only in the CMZ, where such features were mostly known \citep[e.g., ][]{Yusef-Zadeh1984,larosa2001}, but also abundant in central regions relatively far away from the Galactic plane. Some of the radio filaments, observed to be strongly polarized, are clearly synchrotron in origin and are sometimes called nonthermal radio filaments (NTFs). The formation mechanism of such NTFs  remains is still greatly uncertain  \citep[e.g., ][]{Heywood2019,Yusef-Zadeh2019,Sofue2020}.

To understand the nature of these interstellar structures and their connection to Galaxy-wide energetic phenomena, as well as the activity in the GC, we have constructed 
a large-scale high-resolution X-ray map of the GC and its interface with the Galactic bulge, based on  \chandra/ACIS-I observations. They include new observations to fill various gaps in the coverage of 
archival ones. With a total 5.6~Ms exposure of the included observations, this combined mapping, covering a $\sim 2^\circ \times 4^\circ $ field (Fig.~\ref{f:f1}A), now allows us to detect discrete X-ray sources, study their population, distribution, and other properties; map diffuse hot plasma with minimal confusion; and  cross-correlate multi-wavelength objects/features. We here report initial results of the mapping, focused on diffuse X-ray-emitting features and their connection to radio structures in the GC/bulge interface. 

The rest of the paper is organized as follows. In \S~\ref{s:obs}, we briefly describe our data reduction and analysis  procedure. We present our results in \S~\ref{s:res} and discuss their implications for understanding the nature and origin of various interstellar structures in \S~\ref{s:dis}. In \S~\ref{s:sum}, we summarize our main conclusions, as well as results.
In this paper, we adopt a distance of the GC to be  8 kpc (1$^{\prime\prime}$=0.039 pc) and present error bars at the 90\% confidence level.

\section{Observations and Data Analysis}\label{s:obs}

We use all available \chandra\ observations taken before 2020 and with the ACIS-I at the focus. 
They are reprocessed with the standard pipeline of the \chandra\ Interactive Analysis of Observations (CIAO; version 4.11 and CALDB 4.8.4.1). Additional data reduction steps include the subtraction of  the non-X-ray contribution to the data, using the ACIS-I non-X-ray (stowed) background database, the detection of discrete sources and their excision from the data \citep{Wang2004}. Details of these steps are not important for the present work and will be given in a later paper.  

We conduct spatial and spectral analyses of individual X-ray features, as well as point-like sources detected in the survey. In the present work, we focus on the analysis of an X-ray thread \xs\ 
(Fig.~\ref{f:f2}), which is probably the most outstanding vertical linear X-ray feature detected in the \chandra\ survey. The presence of this thread was first reported in an \xmm\ survey of the GC \citep{Ponti2015}. But because of both the limited exposure and spatial resolution of this survey, little specifics can be learned of the thread. In the present \chandra\ survey, the thread is covered by three observations (Obs \# 7157, 19448 and 20111) with a total exposure of 72~ks, all near the axis. We conduct both spatial and spectral analyses of \xs. A fit to the count distribution in the thread give a position angle of 2.7$^\circ$ west from the Galactic north.  The spectral analysis of \xs, in particularly, is based on the on-thread data extracted from a 15$^{\prime\prime} \times 154^{\prime\prime}$ box and a local background estimated in the 100$^{\prime\prime} \times 210^{\prime\prime}$ box east of the thread. Spectral fitting results obtained are not sensitive to the exact choice of these boxes.

\section{Results}\label{s:res}

\begin{figure*} 
\centerline{
\includegraphics[width=1\textwidth]{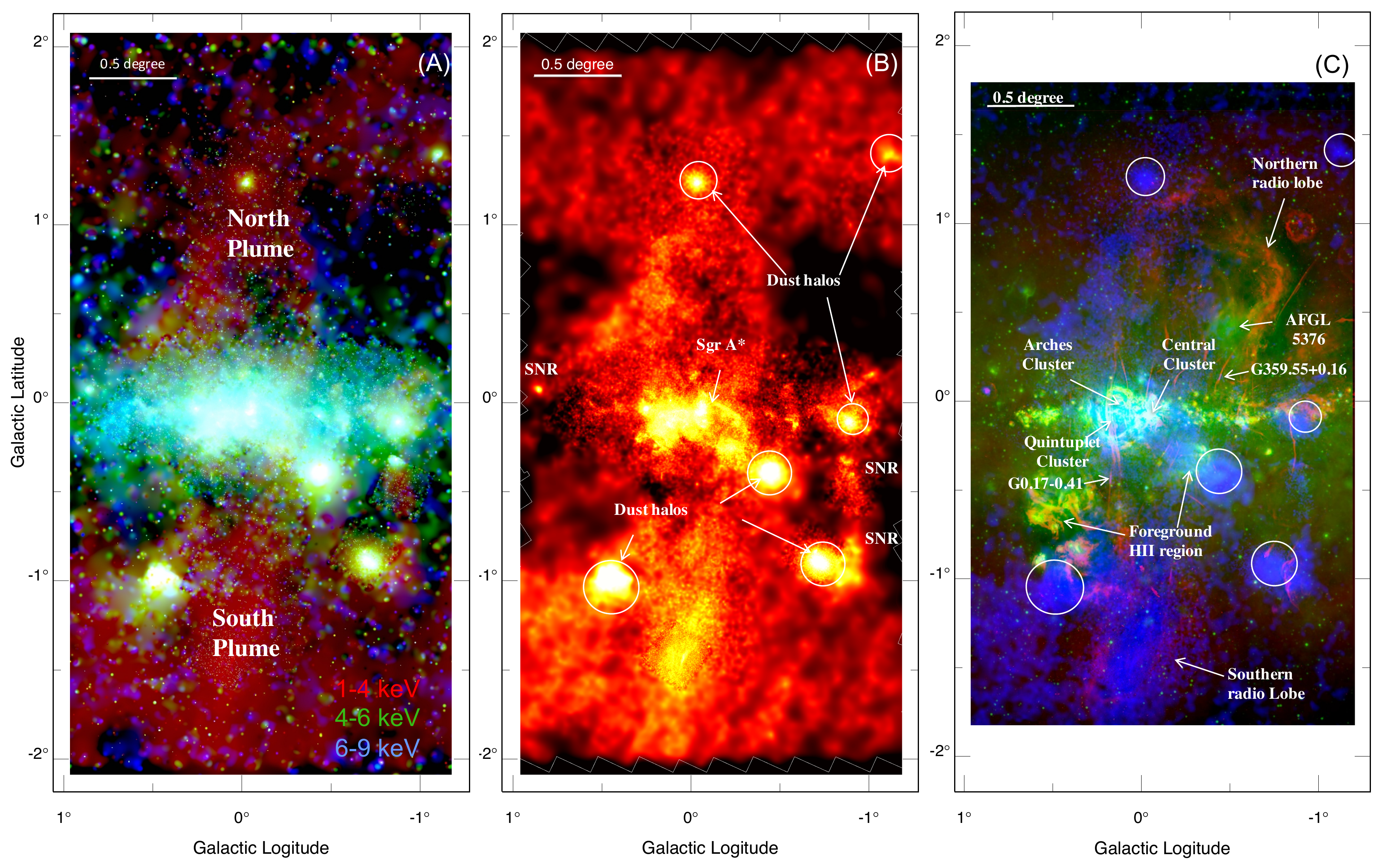}
}
\caption{
\small
\chandra/ACIS-I survey of the GC/bulge field and  a multi-wavelength montage: {\bf (A)} 3-color X-ray intensity mapping -- 1-4~keV (red), 4-6~keV (green) and 6-9~keV (blue); {\bf (B)}  diffuse 1-4 keV map, where detected discrete sources have been excised, although dust scattering halos of bright X-ray binaries remain; 
{\bf (C)}  comparison of the diffuse 1-4~keV map (blue) with the  \meerkat\ 1.3~GHz (red,~\citealt{Heywood2019}) and \wise\ 22\micron\ (green) images. 
 All these plots are projected in the Galactic coordinates. Several major X-ray-bright supernova remnants (SNRs) are marked in (B), while additional key stellar and interstellar structures referred to in the text are labeled in (C). 
}
\label{f:f1}
\end{figure*}

We detect about $10^{4}$ sources in the survey field with the local false detection probability $P <  10^{-6}$. The detection limit depends on the source spectral shape, as well as the local background intensity, exposure, and point spread function, which vary greatly across the field. Nevertheless, we should typically detect sources with luminosities  $\gtrsim 1 \times 10^{32} {\rm~ergs~s^{-1}}$, which include
 black hole/neutron star X-ray binaries, active or quiescent, young pulsars and their associated compact wind nebulae \citep[e.g.,][]{Wang2002,Muno2009,Hong2009,Johnson2009,Jonker2014}. Fainter undetected sources are numerous stars, primarily cataclysmic variables and active binaries (CV+AB; \citealt{Wang2002,Revnivtsev2009}). Their collective contribution should smoothly follow the old stellar distribution in the field and may dominate the ``diffuse'' (detected source-excised) emission above $\sim 4$~keV (more on this topic in \S~\ref{ss:dis-rec-heating}). Here we focus on X-ray-emitting interstellar structures.
 
Figs.~\ref{f:f1}A and B present an overview of the \chandra\  survey of the Galactic center/bulge field. While detailed \chandra\  maps of the generally X-ray bright region along the Galactic plane have been presented previously \citep[e.g.,][]{Wang2002,Muno2009}, our presentation here is optimized to show large-scale features. In this presentation, two outstanding plumes from the GC stand out, mostly in the 1-4~keV band. These plumes, detected previously in \suzaku\ and \xmm\ observations~\citep{Nakashima2013,Nakashima2019,Ponti2019},  are now covered fully at the arcsecond resolution of \chandra\ and with little confusion with discrete sources. The appearance of the plumes is affected by foreground soft X-ray absorption, however, which is most severe at $|b| \lesssim 0.3^\circ$. Here the presence of the plumes is also evidenced by enhanced diffuse 4-6~keV emission, which extends further to higher Galactic latitudes, even in the inner region of the north plume right above Sgr A*.  Foreground absorption remains significant in this part of the plume. Accounting for the absorption effect, the X-ray emission associated with the plumes qualitatively decreases in its intensity and softens in its spectrum with the increasing distance away from the Galactic plane. Therefore, the plumes apparently link the high-energy activity within the CMZ to larger-scale diffuse soft X-ray structures seen at $|b| \gtrsim 1^\circ$ (Fig.~\ref{f:f1}B), where the X-ray absorption is weak. Compared to the north plume, the south plume is substantially more collimated and centrally filled in soft X-ray and extends further away from the GC into the Galactic bulge.

Fig.~\ref{f:f1}C presents a multi-wavelength comparison of various relevant structures seen in the GC/bulge field covered by the \meerkat\ survey~\citep{Heywood2019}. The southern radio lobe, newly discovered by the survey,  encloses the south X-ray plume well. In contrast, the radio and X-ray structures on the Galactic north side are not correlated. The north X-ray plume is systematically offset to the Galactic east from the northern radio lobe, which is sometimes called Galactic Center Lobe  (GCL; e.g., \citealt{Law2009}). The GCL appears to be closely linked to a distinct warm dust  ``double helix'' nebula \citep{Morris2006b}, seen here in the 22 \micron\ emission. The implications of these findings are discussed in \S~\ref{s:dis-rec-ISM}.

\begin{figure} 
\centerline{
\includegraphics[width=0.45\textwidth]{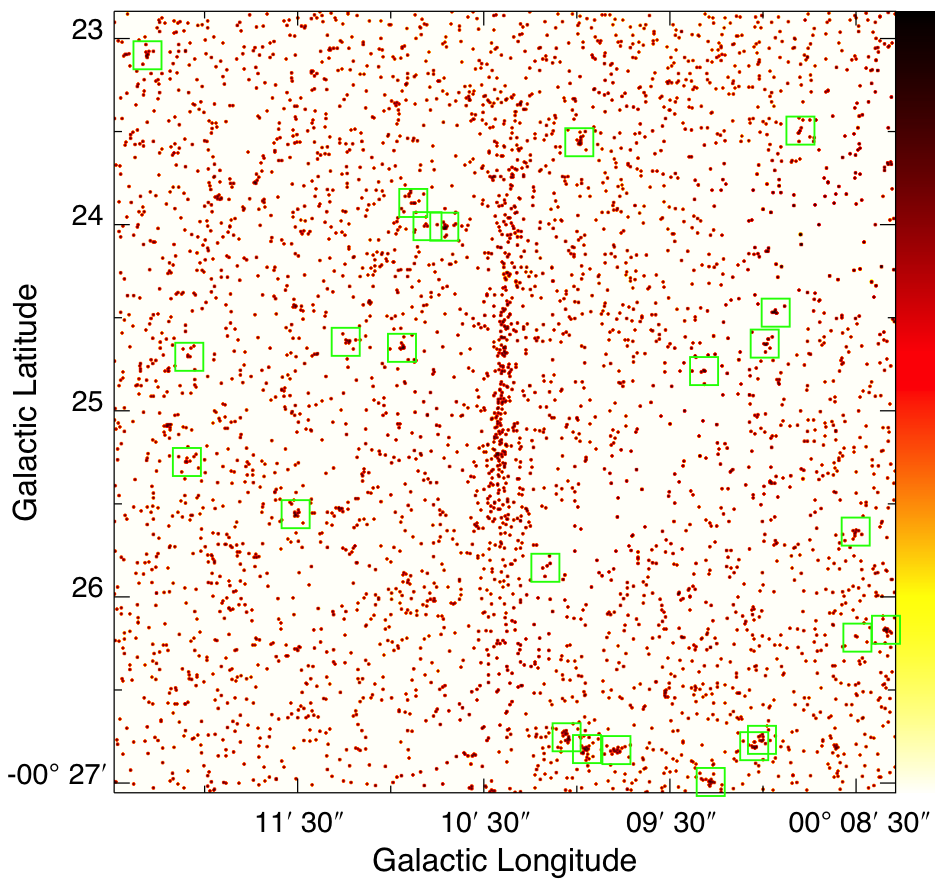}
}
\caption{
\small
High-resolution close-up of \xs\  via  the \chandra/ACIS-I 2.5-6~keV event intensity distribution. Detected discrete sources are marked with squares.
}
\label{f:f2}
\end{figure}

\begin{figure*} 
\centerline{
\includegraphics[width=0.8\textwidth]{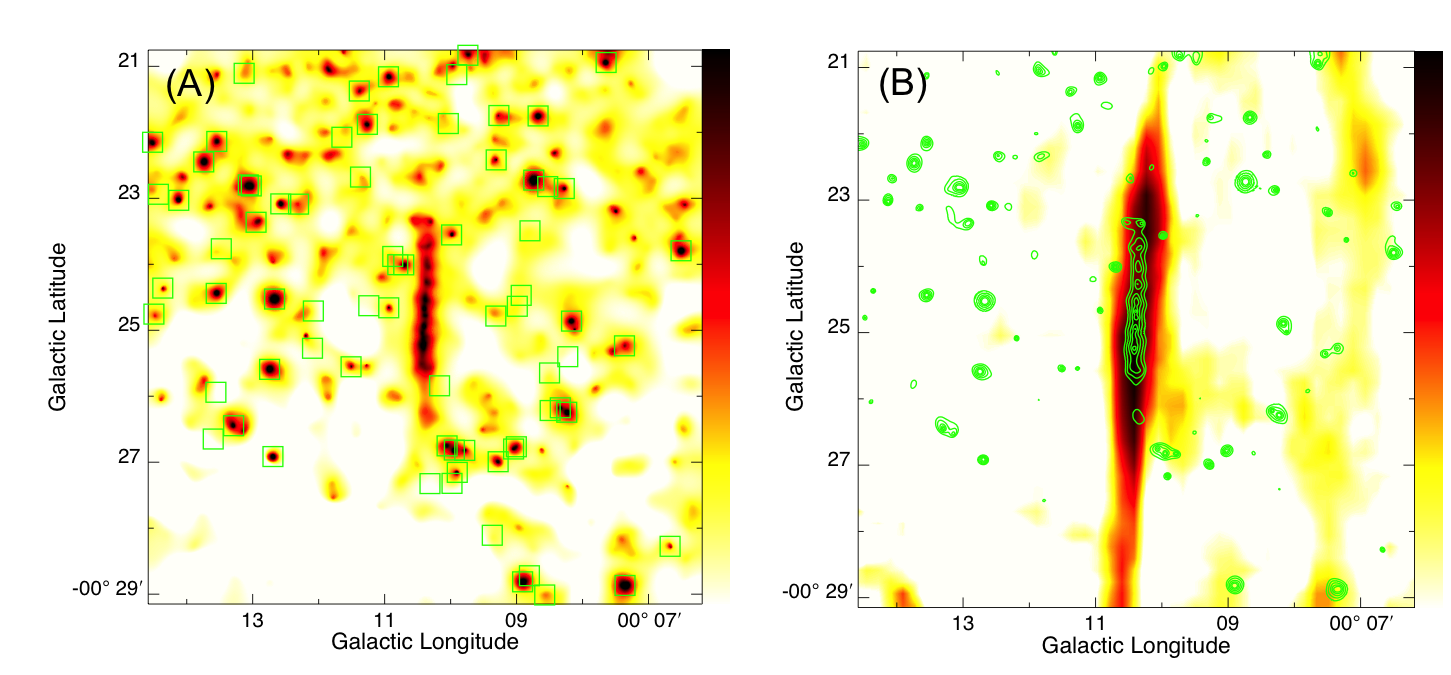}
}
\caption{
\small
Adaptively smoothed \chandra/ACIS-I 2.5-6~keV intensity image of the \xs\  {\bf (A)} and comparison with the \meerkat\ 1.3~GHz image {\bf (B)}, where the X-ray intensity contours  are at 3, 5, 9, 17 and 33 $\sigma$ above the background. 
}
\label{f:f3}
\end{figure*}

\begin{figure*} 
\centerline{
\includegraphics[width=0.8\textwidth]{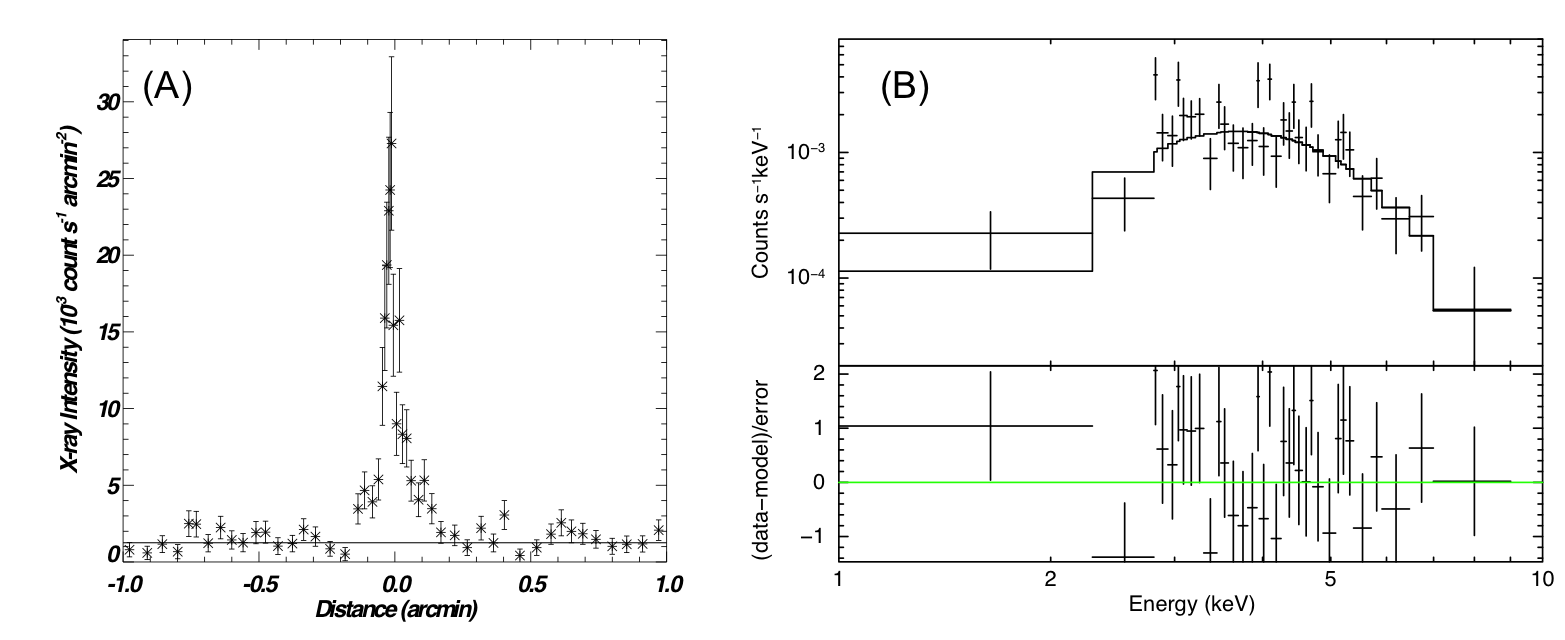}
}
\caption{
\small
 {\bf (A)}  Diffuse 2.5-6~keV intensity, averaged over the length of \xs, as a function of the vertical distance from it; the positive direction is toward the Galactic (slightly south) west. The horizontal line marks the average off-\xs\ background intensity. {\bf (B)}  ACIS-I spectrum of \xs, together with the best-fit power law model (Table~\ref{t:spec}).
}
\label{f:f4}
\end{figure*}

Various small-scale X-ray-emitting linear features or threads are present in the 
\chandra\ data, especially in the inner GC region \citep[e.g.,][]{Wang2002a,Lu2003,Johnson2009,Zhang2020}. Some of these X-ray threads are associated with NTFs. While a systematic identification and analysis of such associations is ongoing, Figs.~\ref{f:f2}-\ref{f:f4} highlight an outstanding X-ray thread, \xs, which is located quite off the CMZ and is nearly vertically oriented, with respect to the Galactic plane. Most interestingly, \xs\ is embedded well within a NTF, which appears to be locally swelled (Fig.~\ref{f:f3}B). Their physical association is thus evident.

The presence of \xs\ is most significant in the 2.5-6~keV band and does not seem to be related to any 
detected discrete sources, as shown in Fig.~\ref{f:f2}, which is nearly at the full resolution of \chandra. 
The thread is very narrow in its middle part and becomes dimer and wider on both sides. The image presented in Fig.~\ref{f:f3} is adaptively smoothed to reach a non-X-ray background-removed signal-to-noise ratio of $> 4$ across a larger field of view that is linearly twice larger than that covered by Fig.~\ref{f:f2}. The total length of the thread is about 2\farcm6 or $l_{x} \approx 6$~pc at the distance of the GC.  In both Figs.~\ref{f:f2} and ~\ref{f:f3}, \xs\ shows substructures, although individually they are only marginally significant at $\lesssim 2\sigma$ levels.   Fig.~\ref{f:f4}A shows a surface brightness profile from a cut perpendicular to the full length of  \xs\ and is adaptively grouped to achieve $S/N > 5$ for each bin. We use a similar profile for a cut through the inner half of the length to estimate the effective width of the thread as $\sim 1\farcs5$ (FWHM).

\begin{table} 
{\centering
\caption{X-ray spectral fitting results of \xs}\label{t:spec}
\medskip
\begin{tabular}{lrrr}
\hline\hline
Parameters &  Power law & \multicolumn{2}{c}{APEC}\\
 \hline
N$_{\rm H} (10^{22}{\rm~cm^{-2}})$&11 (5.4-17) & 5.2 (3.8-7.0)& 9.0(5.1-15)\\
$\Gamma$ or $kT$ (keV)&3.0 (1.6-4.5) & $>7.9$& 3.2(1.5-14)\\
$Z$ (solar abundance) & - & 1& $< 0.51$\\                    
$K$  & $5.7 \times 10^{-4}$ & $1.5 \times 10^{-4}$& $4.9 \times 10^{-4}$\\
$\chi^2/{\rm d.o.f}$&36.0/32 &39.4/32& 35.8/31 \\
$f_{x} (10^{-13}{\rm~erg~s^{-1}~cm^{-2}})$&1.3 &1.6& 1.2 \\
$L_{x}(10^{33}{\rm~erg~s^{-1}})$& 2.6 & 1.8& 2.1\\
\hline
\end{tabular}
}

Note: The parameter uncertain ranges are all presented at the 90\% confidence level. $K$ is the model normalization. For the APEC modeling, two fits are conducted: one assumes that the metal abundance is solar, and the other lets it to be fitted. For each model fit, both flux $f_{x}$ and luminosity $L_{x}$ are inferred from the best-fit model in the 2-10~keV band. 
 $L_{x}$ further corrects for the absorption (i.e., N$_{\rm H} =0$). 
\end{table}

With a net count rate of $(5.0\pm0.4) \times 10^{-3} {\rm~cts~s^{-1}}$, the ACIS-I detection of \xs\  allows for simple characterizations of its spectral shape. Table~\ref{t:spec} lists model parameters: a power law and an optically-thin thermal (APEC), assuming collisional ionization equilibrium.  Solar metal abundances~\citep{Anders1989} are assumed for both the hot plasma. Fig~\ref{f:f4}B shows the power law fit to the X-ray spectrum of this thread, which is only slightly more favored statistically than the plasma characterization with the abundances fixed to the solar values (first column under APEC in Table~\ref{t:spec}). The APEC model fit can be marginally improved, if the abundance is allowed to vary, which suggests an abundance of $< 0.51$ solar (at the 95\% confidence; second column under APEC in Table~\ref{t:spec}). This abundance is somewhat lower than what is expected for the GC, but is consistent with the value obtained for \rxs\  \citep[][see \S~\ref{s:dis-cont} for more discussion]{Yamauchi2014}. The large X-ray-absorbing column density N$_{\rm H}$ inferred from either model fits is consistent with the GC location of \xs. 

Adopting the optically thin thermal plasma characterization of the X-ray emission from \xs, we can estimate the density as $n \approx n_e \approx 20 D_{kpc}(K/V)^{0.5} f_h^{-0.5}  {\rm~cm^{-3}} \approx 15 f_h^{-0.5} {\rm~cm^{-3}}$, where $D_{kpc} \approx 8$ is the distance to \xs, $K$ is from Table~\ref{t:spec}, $V \approx 0.017 {\rm~pc^{3}}$ is the total volume, assuming a cylinder of diameter $\approx 0.06$~pc (roughly the FWHM of the thread) and the above length estimate $l_x $, and  the effective volume filling factor $f_h$ of the hot plasma is probably least certain. The total thermal energy and the radiative cooling time are $E_{th} \sim 3 n_e kT f_{h}V\sim 3 \times 10^{47} f_{h}^{0.5}$~erg and $t_c \sim 5 \times 10^{7} f_h^{0.5}$~yr, where $kT \sim 10$~keV is assumed. The dynamic age of the thread may be approximated as the sound-crossing time of the re-connection region: $t_d \sim l_x/c_s \sim 6 \times 10^{3} (kT/10{\rm~keV})^{-0.5}$~yr, over which the radiative cooling is negligible.  A similar age may also be obtained if the Alfv\'en speed, $2.2 \times 10^{3} n^{-0.5} B_{\rm mG} {\rm~km~s^{-1}}$, is used.

\section{Discussion}\label{s:dis}

Here we first compare \xs\ with other most relevant X-ray threads and NTFs observed in the GC, providing both a brief review of existing work and a context for later discussion, then explore potential origins of \xs\ and another similar X-ray thread/NTF association \rxs, and finally focus on the potential role of magnetic field re-connection events in regulating the ISM in the GC and beyond.

\subsection{Comparison with other similar X-ray and/or radio features}\label{s:dis-cont}

Among the X-ray/NTR associations observed in the GC, \xs\ most resembles \rxs. Discovered by \citet{Lu2003} in an early \chandra\ GC survey, \rxs\ also shows a linear X-ray structure of a length similar to \xs's, contains no point-like source, and is located quite far away from and oriented nearly perpendicular to the Galactic plane. The associated NTF represents a strand of flux tubes, consisting of two main parallel bundles. Radio polarization measurements show that the average magnetic field is orientated along these bundles \citep{Yusef-Zadeh1997}. With a deep exposure of \suzaku, \citet{Yamauchi2014} show that the X-ray spectrum of \rxs\ exhibits a distinct FeXXV-He$\alpha$ 6.7-keV emission line  with an equivalent width of $850^{+400}_{-390}$ eV. The spectrum can be well characterized by an optically-thin thermal  plasma with the temperature and metal abundance of  $4.1^{+2.7}_{-1.8}$ keV and $0.58^{+0.41}_{-0.32}$ solar, which are qualitatively similar to those obtained for \xs\ (\S~\ref{s:res}). Because of these similarities between \xs\ and \rxs, we explore their origins together in the following. To do so, we also need to place them in a large context of  the NTFs observed in the GC.

The NTFs, although most of which do not have any X-ray counterparts, share some common morphological characteristics, as well as radio spectral and polarization properties. One of the striking collective characteristics of the NTFs is their predominantly radial orientation and concentration in the GC. They appear mostly radiating from within $\sim 80$~pc of Sgr A*, tracing the poloidal magnetic field component or the magnetosphere of the GC \citep{Morris2006a}. The component may represent protogalactic field that has been concentrated over the history of the Galaxy by mass inflow~\citep{Chandran2000} and/or outflows themselves under the influence of the Coriolis force and magnetic re-connection \citep[e.g.,][]{Hanasz2002}.  Analyses of existing observations show that the NTFs likely represent the periphery of the ongoing outflows from the GC~\citep{Heywood2019,Yusef-Zadeh2019}.  Incidentally,  \xs\  is at the east rim of the southern radio lobe, while \rxs\ is at the western boundary of the north plume of X-ray emission. \rxs\ also seems to connect between the CMZ and a warm dust emission region, AFGL 5376  (Fig.~\ref{f:f1}C; \citealt{Uchida1994}). Indeed, the NTFs all appear to be undergoing an interaction with a molecular cloud \citep[e.g.,][]{Morris2015}.  

The most tight constraint on the magnetic field strength inside NTFs relies on a simple dynamical argument, originally proposed by \citet{Yusef-Zadeh1987b}.  They argue that the absence of any significant bending or distortion of an NTF against the strong ram pressure of its associated molecular cloud indicates a field strength of $\gtrsim 1$~mG.  The fattening of the NTF associated with \xs\  suggests an over-pressure produced by the X-ray thread. Adopting the optically thin thermal plasma characterization of this X-ray emission from \xs\ (\S~\ref{s:res}) and assuming the balance between the outward thermal pressure and inward magnetic tension, we estimate the required total vertical field strength $B_{\perp} \sim 3 f_h^{-0.25} $~mG.  This value is quite large and may not be provided by the fattening of the NTF alone. Other contributions to the confinement can also be important. We will come back to this point in \S~\ref{ss:dis-sc-rec}.

The above discussion, together with our new results, now provide a context for us to explore the origins of such X-ray thread/NTF associations as \xs\ and \rxs.

\subsection{Pulsar wind nebula interpretation }\label{ss:dis-sc-pwn}

Based on the satisfactory power law characterization of the X-ray spectrum (Table~\ref{t:spec}), one may suggest that \xs\ could represent a pulsar wind nebula (PWN), confined by a strongly magnetized  medium \citep{Wang2002a,Bandiera2008,Barkov2019a,Barkov2019b}. In this scenario, the X-ray thread would be interpreted as the synchrotron emission of relativistic electrons (and positrons), while streaming mostly along magnetic lines. Indeed, the scenario has been proposed for multiple X-ray threads in the GC, with or without radio counterparts \citep{Wang2002a,Johnson2009,Zhang2020}, although none of these cases has been confirmed by the detection of a putative pulsar. The scenario is considered to be most plausible when an X-ray thread contains a point-like X-ray source (i.e., the putative pulsar) and/or exhibits evidence for synchrotron cooling  \citep{Wang2002a,Gaensler2004,Park2005,Wang2006,Zhang2020}.  

One of -- if not the -- best candidates for such a PWN interpretation in the GC is G0.13-0.11, which
shows a distinct curved linear X-ray morphology  and borders a bow-shaped radio feature
\citep{Wang2002a,Zhang2020}. The properties of the X-ray and radio emissions and their  morphological configuration strongly suggest that G0.13-0.11 results from a moving pulsar with the propagation of its wind material  strongly regulated by surrounding organized magnetic field \citep{Wang2002a,Zhang2020}. 
Well embedded in the X-ray thread is a point-like source of $L_{2-10{\rm~keV}} =  8.4  \times 10^{32}{\rm~erg~s^{-1}}$. The X-ray properties of both the source and the thread are consistent with the PWN interpretation  \citep{Wang2002a,Zhang2020}.  G0.13-0.11 is also apparently associated with a point-like TeV source, detected at a significance of 5.9$\sigma$  in the HESS GC survey~\citep{Aharonian2006,H.E.S.S.Collaboration2018}. The inverse-Compton origin of the TeV emission matches well with the synchrotron interpretation for the X-ray emission of G0.13-0.11. It may thus be considered as a reference source for us to evaluate the likelihood for \xs\ or \rxs\ to be a PWN.

We find that an application of the PWN (or SNR) scenario to \xs\ or \rxs\ meets several major difficulties. 
\begin{itemize}
\item There is no evidence for a point-like source as the putative pulsar in either X-ray thread. Assuming the X-ray versus spin-down luminosity ($L_{sd}$) correlation, log$L_{2-10{\rm~keV}} = 1.34 {\rm~log} L_{sd} - 15.34$ \citep{Possenti2002}, one would expect a quite powerful pulsar with $L_{sd} \sim 3 \times 10^{36}{\rm~erg~s^{-1}}$. Such a source, if present in \xs\ for example, would be easily detected, because its overall X-ray luminosity is about the same as that of G0.13-0.11. 

\item The power law photon index of \xs\  (though with a large statistical uncertainty range; Table~\ref{t:spec}) is not typical for a PWN and is steeper than that of G0.13-0.11, for example, at the 95\% confidence level. For \rxs,  the PWN scenario for its X-ray emission can be completely ruled out because the detection of the strong He-like Fe 6.7-keV emission line, which arises from an optically-thin thermal plasma. 

\item  The PWN scenario also has the difficulty to explain the X-ray thread length of \xs. The X-ray synchrotron cooling time scale is $t_{s,x} \sim 1 {\rm~yr}~B_{mG}^{-1.5}~E_{keV}^{-0.5}$, where $ E_{keV}$ is  photon energy. Thus $t_{s,x}$ is too small to be compared with the time scale for the electrons to move cross the thread, which suggests that the particle acceleration cannot be in a point-like source, unless $B_{mG}$ is substantially less than 1, which is not favored for such a bright NTF. 

\item Because of the long radio-synchrotron cooling time scale, one may expect that a pulsar with a typical moving speed (a few $\times 10^{2} {\rm~km~s^{-1}}$) should leave an extended radio remnant or PWN of a fan-out morphology extending only on one side of the X-ray thread and at the opposite direction of the pulsar's proper motion, as is the case for G0.13-0.11, but not for \xs. The fattening of the NTF around \xs\ is rather symmetric and has a half width of $\lesssim 1$~pc, as seen in the 1.3 GHz image. If the radio synchrotron cooling dominates, then we may constrain the vertical speed of the pulsar crossing the NTF filament to be $\sim 5 {\rm~km~s^{-1}} B_{mG}^{3/2}$. This speed is much smaller than the velocity expected for a pulsar or even for a source with a relative orbital motion, which is common in the GC. Clearly, a small $B_{mG}$ value would make this problem worse. Thus it is difficult to understand such a nearly perfect and symmetric alignment between the X-ray thread and the  NTF, unless they are physically associated and are produced by the flux tube itself. A similar argument can also be made for \rxs. 

\item There is no indication of any TeV emission enhancement at the location of \xs\ or \rxs\ in the HESS GC survey~\citep{Aharonian2006}. The diffuse TeV background at the locations of these threads, away from the CMZ, is much low, in comparison to the location of G0.13-0.11. The lack of the HESS detection indicates that the flux of a point-like source would be at least about 30 lower than that of G0.13-0.11. This limiting TeV flux needs to be explained if the X-ray flux  of \xs, for example,  is to be interpreted as synchrotron emission.

\end{itemize}

Finally, could \xs\ and \rxs\ be SNRs regulated by their strongly magnetized environments?   Both, more so \xs, are located quite off the CMZ, where recent star formation occurs. Correspondingly, the probability should be small to detect a young pulsar or an SNR at such a location, especially in coincidence with a probably distinct flux tube of strongly enhanced magnetic field. Moreover, our estimated thermal energy of \xs\ (\S~\ref{s:res}; similar for \rxs) is far too small to be consistent with that expected for a typical SNR. Its age should not matter much here, because the thermal radiative cooling time is long. This energy estimate should also hardly depend on the thermal emission assumption. Fundamentally the pressure and volume of the X-ray thread are too limited to be consistent with an SNR interpretation. However, it may be possible that \xs\ or \rxs\  represents just a small piece of a SNR, ejected from the CMZ along a magnetic flux tube, although the plausibility of this scenario is yet to be carefully examined, both theoretically and observationally.

\subsection{Magnetic field re-connection scenario }\label{ss:dis-sc-rec}

This scenario provides a unified explanation for the X-ray thread/NTF associations of
\xs\  and \rxs, in terms of their formation and emission. We consider that they represent interstellar magnetic field strands. Various models have been proposed for the formation of such structures, mostly related to the outflow or wind from the GC \citep[e.g.,][]{LaRosa2004, Sofue2005,Banda-Barragan2016, Yusef-Zadeh2019}. 
 
 In the cometary model \citep[e.g.,][]{Shore1999}, for example, NTFs are magnetized wakes generated by the interaction of molecular clouds with a global GC wind. Naturally, field lines of opposite directions are expected to meet at the rear of such clouds, forming re-connection-prone wakes -- a situation similar to what happens in the heliospheric tail \citep[e.g.,][]{Lazarian2010,Banda-Barragan2016}. The lines could also be twisted, e.g., due to the spin of the molecular clouds and/or the different rotation of the CMZ~\citep{Mezger1996,Yusef-Zadeh1987a,LaRosa2004}. The twisting leads to the formation of the tightened ropes of enhanced magnetic field flux tubes and eventually to the magnetic re-connection \citep[e.g.,][]{Titov1999,Tsap2020}.

Re-connection (more precisely breaking and rejoining) events take place at interface between ionized or partially ionized gases with opposite magnetic field lines  \citep[e.g.,][]{Lazarian1999,Zweibel2009,Bicknell2001,Lazarian2020}. In such an event,  the fractal tearing instability of field lines leads to the formation of plasmoids (magnetic islands) in the re-connection region \citep[e.g.,][]{Furth1963}. 
Both observations and simulations have shown that the local vertical field strength in re-connection regions become comparable to or even slightly greater than the large-scale mean field \citep[e.g.,][]{Ieda1998,Lu2020}. The total pressure inside a plasmoid is typically higher than that in the background because magnetic tension force acts to compress the plasmoid. The situation here is analogous to the sun, where the energy density in solar flare regions can be much stronger than the average value in the solar corona~\citep[e.g.,][]{Toriumi2019}. Magnetic re-connection is also a dynamic process. 
Electric field generated by the changing magnetic field and the collision among the plasmoids further accelerate particles, producing cosmic-rays (CRs; e.g., \citealt{Bicknell2001}). Direct observational evidence for re-connection events is abundant in the solar corona and space weather~\citep[e.g.,][]{Toriumi2019}, but so far little in the ISM. 

The production of diffuse hot plasma via re-connection in the interstellar space has been predicted theoretically, in analogy to the heating of the solar corona \citep[e.g.,][]{Heyvaerts1988,Raymond1992,Tanuma2003,Florido-Llinas2020}, and has been used to explain diffuse X-ray emission observed in nearby galaxies \citep[e.g.,][]{Wezgowiec2020}, as well as in our Galaxy  \citep[e.g.,][]{Tanuma1999}. MHD simulations in interstellar contexts \citep{Tanuma1999,Tanuma2003,Hanasz2002} show that re-connection events can occur in expanding shells (e.g., due to Parker or magnetic buoyancy instability) and, under the influence of Coriolis force,  produce ``helical tubes'' of  magnetic field lines, as well as in wakes of molecular clouds interacting with a global GC wind \citep[e.g.,][]{Shore1999, Banda-Barragan2016}. This mechanism seems to naturally explain the close X-ray/NTF associations, as seen in \xs\ and \rxs, as well as their morphological and spectral properties.  The 6.7-keV iron line emission, in particular, is also most naturally explained by an interstellar re-connection event \citep{Lu2003}. \xs\ probably represents an even stronger case than \rxs\ as a re-connection event in an interstellar magnetic field rope. Much of the X-ray emission may arise from plasmoids as indicated by the discrete substructures of \xs\ (\S~\ref{s:res}). The small-scale radially inward tension force of magnetic loops around these plasmoids may largely responsible for the confinement of  \xs\ and \rxs. This small-scale plasmoid confinement is in addition to the global inward tension of the curved field lines of the NTF around \xs, as discussed at the end of \S~\ref{s:dis-cont}. 

 Furthermore, a reconnection event is likely triggered by external overpressure of the ambient medium  (e.g., due to the cloud-cloud collision and/or twisting of a magnetic rope), although no direct evidence is available for such a dynamical process around the \xs/NTF association. Our argument for the reconnection scenario is that it provides a consistent interpretation for the X-ray properties of \xs\  and \rxs, as well as their physical association with the NTFs. While the X-ray threads likely represent the ongoing or recent reconnection sites, the associated NTFs are the illuminated parts of the underlying magnetic flux tubes, which could extend to much larger scales and even connect to features on the opposite side of the Galactic plane (e.g., Fig.~\ref{f:f1}C). Therefore, the X-ray thread/NTF associations such as \xs\ and \rxs\ may provide potentially excellent laboratories to test our understanding of the re-connection astrophysics in the ISM. They allow for spatially resolved studies of the hot plasma production, as well as the particle acceleration and magnetic field structure, in the re-connection zone. 

In the re-connection scenario, the X-ray emission directly traces the heated thermal plasma.  According to \citet{Tanuma2003}, the plasma is heated to a characteristic temperature of $kT \sim 10 n^{-1}B_{\rm mG}^{2}$~keV. The density $n$ of the gas accompanied with the magnetic field can vary greatly from one location to another in the GC. But  most ubiqious is diffuse warm molecular gas of  $n \lesssim 100 {\rm~cm^{-3}}$\citep{Oka2019} or  more tenuous diffuse hot plasma. 
The estimated  $n$  and $B$ values (\S~\ref{s:res} and S~\ref{s:dis-cont})  are reasonably consistent with those required to produce $kT \sim 10$~keV plasma via the re-connection. Using the results given in \S~\ref{s:res}, we also infer the plasma heating rate of $\sim E_{th}/t_d \sim 2 \times 10^{36} {\rm~erg~s^{-1}}$.  The total energy generation rate of the region should be larger than this value, accounting for the energy in the plasma motion and in the particle acceleration. 

Different radio/X-ray associations may represent re-connection events at different evolutionary stages \citep{Tanuma1999,Tanuma2003,Hanasz2002}. The swelled radio morphology of \xs\  indicates that the X-ray-emitting material is dynamically important and is still shaping the surrounding magnetic field topologically, which may represent an early event stage (e.g., Fig. 6 in \citealt{Tanuma1999}).  
In comparison, \rxs\  may be a manifest of a later-stage re-connection event.
The NTF morphology around this thread is topologically different from that associated with \xs. \rxs\ appears at the closest (possibly crossing) part of the two major bundles of the NTF.  Off the X-ray thread, they branch out.  The X-ray thread is associated with only one of them at present.  There is no evidence for any significant local inflation of the associated NTF in \rxs, suggesting that magnetic field tension dominates the hot plasma pressure. Some of  the plasma may have already escaped from the re-connection region, at least from parts of the NTF. Therefore, the re-connection scenario provides a consistent explanation of the X-ray properties of \xs\ and \rxs, as well as their association with the distinct NTFs. 

Of course, more conclusive evidence will be needed to establish the re-connection scenario.  
Future more sensitive X-ray and radio observations will enable spatially-resolved and even time-dependent spectroscopy of such re-connection phenomena.  
In particular, a deeper X-ray observation of \xs\ will better constrain the presence of of the 6.7-keV line, as expected for the re-connection scenario. Multi-band high-resolution radio observations with polarization measurements will be extremely useful to explore the magnetic field structure, as well as the propagation properties of the CR electrons along the associated NTFs. Dedicated theoretical studies, including computational simulations, are needed to improve our understanding of the magnetic re-connection in various interstellar environments.
For now, we can examine potential connections of  re-connection events to other NTFs and to larger-scale interstellar structures observed in the GC.

In addition to \xs\ and \rxs, some of the other NTFs in the GC may also arise from re-connection events. Their X-ray emission may just be too weak and/or diffuse to be detected individually. As shown in MHD simulations \citep{Tanuma2003}, the energy release profile of a re-connection event is strongly model-dependent and typically shows an overall fast-rising slow-decay pattern.  The half-peak duration is on the order of 10 times the sound-crossing time $t_{d}$. Many re-connection events may not be able to heat plasma to high enough temperature for significant X-ray emission. Identified radio/X-ray filament associations likely represent only the tip of the ``re-connection iceberg''.
Its true magnitude  may be better traced by NTFs.  CR electrons diffuse along the field lines at the Alfv\'en speed over the characteristic synchrotron cooling time scale $t_{s,r}$ \citep{Morris2006a,Thomas2020}. One then expects to detect radio emission off the re-connection sites on the spatial scale of $44 (n B_{\rm mG} \nu_{\rm~GHz})^{-0.5}{\rm~pc}$, which is comparable to the observed length of the longest
filaments in the GC. The resultant NTFs should in general be more extended and  live longer than the X-ray threads, again consistent with the observations.  Therefore, we may reasonably assume that  these re-connection events are intimately linked to energetic outflows as traced by  the large-scale interstellar structures in the GC.

\subsection{Re-connection and global ISM structures}\label{s:dis-rec-ISM}

To explore the potential links of magnetic field re-connection to interstellar structures observed in various wavelengths, we first need to determine whether or not and how they may be physically associated in the GC. Here we focus on the X-ray plumes and the radio lobes, as we have described in \S~\ref{s:res}.

\subsubsection{Origin of the X-ray plumes and their connection to the radio lobes}\label{ss:dis-origin-str}

In the GC field,  projection effect can be significant. Particularly relevant here is the GCL. It has recently been proposed to be a foreground HII region \citep{Nagoshi2019,tsuboi2020}. This proposition appears consistent with various observations: the detection of the associated H$\alpha$ emission~\citep{Law2009}, the small systemic velocity gradient (between $-4 $ and $+10 {\rm~km~s^{-1}}$) observed in radio recombination lines \citep{Law2009,Alves2015,Nagoshi2019,Sofue2013}, and the lack of an X-ray counterpart (Fig.~\ref{f:f1}C). Even the brightest (western) part of the lobe shows no enhanced X-ray emission in either 1-4~keV or 4-6~keV band (Fig.~\ref{f:f1}A). While the X-ray plume seems to be rooted in the most active portion of the CMZ, including all the three massive-and-young star clusters, the GCL is offset to the Galactic west and is not associated with any evident energetic source in the GC (Fig.~\ref{f:f1}C).
An ongoing multi-wavelength study is placing tight constraints on the line-of-sight location of this HII region and on a nearby young stellar cluster that may be the source for the ionizing radiation \citep[private communications]{Benjamin2020}.
One may speculate that the HII region is also responsible for much of the angular size and temporal pulse broadening observed in GC radio pulsars, especially the transient magnetar SGR J1745-2900, which is only 2\farcs5  from Sgr A*~\citep{Bower2014,Sicheneder2017}. Indeed, a single HII region modeling of the broadening measurements places the radio scattering screen at 1.5-4.8~kpc away from Earth, probably in the Scutum spiral arm~\citep{Sicheneder2017}. Therefore, we consider the GCL to be a foreground of the GC and is not physically related to the X-ray plumes.
 
The north X-ray plume (Fig.~\ref{f:f1}) does not seem to be enclosed by any coherent radio structure and may represent an aged outflow or fountain of hot plasma, which is apparently powered by some or all of the young massive stellar clusters. Unlike in the Galactic disk, the dynamics of the outflow in the GC may be strongly influenced by the motion of Arches and Quintuplet clusters, which orbit around Sgr A* with period $\sim 10^{6}$~yrs, shorter than their ages.
In contrast, the physical association of the south X-ray plume with the newly discovered \meerkat\ southern lobe \citep{Heywood2019} appears secure. The radio lobe extends to the southern edge of the field in Fig.~\ref{f:f1}C and wraps around the X-ray plume  well. 
The X-ray emission from the plume is clearly dominated by optically-thin thermal plasma in an over-ionized state \citep{Nakashima2013}. This state, together with the strongly centrally-filled and collimated morphology, suggests that the plume represents a recent blowout from  the high-pressure GC region, resulting in the over-cooling of the enclosed plasma. The collimation of the hot plasma and its association with the radio lobe further indicate that the confinement of the plasma is largely due to magnetic field lines, which are likely anchored in the CMZ and are twisted because of its orbital motion. 

\subsubsection{Heating of diffuse hot plasma in the GC}\label{ss:dis-rec-heating}

Diffuse hot plasma  is best traced by X-ray emission. However, there is an ongoing debate as to the nature of the X-ray emission from the GC.
Relevant studies have largely been focused on the nature of the prevalent FeXXV-He$\alpha$ (6.7-keV) line emission \citep[e.g., ][]{Nishiyama2013,Koyama2018}. This line emission has been interpreted mostly as a superposition of unresolved point sources (CV+AB; \citealt{Wang2002}), which is largely confirmed in a 1~Ms deep \chandra\ exposure of an inner bulge field ($l \sim 0.08^\circ, b \sim -1.42^\circ$; \citealt{Revnivtsev2009}). But extending this interpretation to the GC has been  challenged  vigorously \citep[e.g., ][]{Nishiyama2013}. \suzaku\ observations have shown that the specific line emission per stellar mass in regions outside $|b| \sim 0.5$ is consistent with the Galactic 
bulge/ridge value, but increases (by a factor of up to $\sim 5$) with the decreasing $|b|$  in the inner GC region. Metallicity increase toward the GC could be a factor, although the iron abundance there is on average   only up to about twice solar  \citep[e.g.,][]{Nogueras-Lara2018}. Alternatively, the enhanced line emission, as well as the rather flat X-ray spectral continuum \citep[e.g.,][]{Wang2002}, 
may be largely due to the presence of diffuse plasma with a characteristic temperature of  $\sim 7$ keV. This scenario, if confirmed, would have profound implications for our understanding the heating and confining
mechanisms in the GC \citep[e.g.,][]{Nishiyama2013}; a huge amount
of magnetic energy ($\sim 10^{55}$~erg) would be expected in the central $\sim 300$~pc, comparable to the kinetic energy associated with the rotation of the gas in the CMZ \citep{Ponti2015}. There are also lines of evidence for a diffuse soft  X-ray component which is most likely due to plasma with a characteristic temperature probably at 
$\sim 1$~keV \citep[e.g.,][]{Ponti2015,Yamauchi2018}. Such a characterization, however, is sensitive to the spectral model assumed in the multiple-component decomposition of the observed X-ray emission.

The presence of a substantial diffuse hot plasma component in the GC now becomes clear, with the detection of the distinct interstellar structures such as the north and south X-ray plumes, as revealed in \S~\ref{s:res} (see also \citealt{Nakashima2013,Nakashima2019,Ponti2019}). These plumes apparently originate in the most active portion of the CMZ and provide links to global energetic structures in the bulge and halo of our Galaxy.

Much of the diffuse X-ray emission from the plumes and from the GC in general could be produced by re-connection events. The magnetic pinch confinement of super hot plasma in individual flux tubes helps to mitigate the energy supply problem  \citep[e.g.,][]{Nishiyama2013}, as it is not freely escaping from the GC. As discussed in \S~\ref{ss:dis-sc-rec}, most of 
such hot plasma-filled flux tubes may not be detected individually with existing data. 
Because of the long radiative cooling time, the plasma will eventually expand and cool down, converting some of the thermal energy to kinetic energy. Therefore, one may expect that this unresolved plasma contribution should show a softer X-ray spectral shape than that in such confined threads as \xs\ and probably \rxs, consistent with the observed spectral properties of the diffuse X-ray emission from our \chandra\ survey field. Therefore, re-connection events may collectively contribute significantly to the X-ray emission observed in the GC.   

One may also expect that re-connection events should occur more frequently and vigorously in and near the CMZ, where the overall interstellar pressure is more extreme. This is not only because of strong energy feedback from massive stars and possibly Sgr A*, but also due to differential orbital and turbulent motion of magnetized ISM. Indeed, many NTFs have been observed in the CMZ.  Some of them do have X-ray counterparts  \citep[e.g.,][]{Wang2002a,Lu2003,Johnson2009,Zhang2020}, although detection or identification of individual re-connection events is  more difficult because of the confusion with other X-ray-emitting structures and the overall enhanced background. 

\subsubsection{Re-connection-regulated global ISM}\label{ss:dis-rec-ISM}

A key ingredient for the re-connection to be an important interstellar process is the presence of a sufficiently large collective interface where magnetic fields of opposite directions collide. Such a dynamical event can be realized in various ways in a strongly turbulent medium, driven by differential rotation, as well as recent energetic feedback from massive stars and perhaps Sgr A*. High-velocity collisions can occur not only in the CMZ, but also around it. Energetic outflows, as evidenced by the radio lobes and X-ray plumes, naturally stretch mostly toroidal magnetic field lines in the CMZ to high Galactic latitudes, leading to the formation neutral sheets between colliding media with different magnetic field orientations. Strongly enhanced flux tubes may also form out of swept up material, e.g., due to collisions between poloidal magnetic fields and a jet-like outflow from Sgr A* \citep[e.g.,][]{Sofue1987} or dense clouds \citep{Serabyn1994}, stripping of  gas off dense gas clouds \citep[e.g.,][]{Shore1999, Banda-Barragan2016}, and the presence of various other obstacles \citep[e.g., individual stellar wind bubbles][]{Yusef-Zadeh2019}. These processes and hence associated re-connection events tend to occur at the outer boundaries of global outflows.  As pointed out in \S~\ref{s:dis-cont},  the \xs\  and \rxs\ and their associated NTFs indeed appear at the rims of the radio lobes or X-ray plumes. This should motive future observations of gas associated with the southern radio shell, for example, to study its kinematics and potential interplay with various NTFs. Moreover, flux tubes in outflowing plasma can be naturally produced by rotational motion of associated molecular clouds in the GC, as shown in simulations \citep[e.g.,][]{Mezger1996}.  In fact, recent 3-D simulations show that turbulence and re-connection in the ISM are inter-connected on all scales: re-connection is an intrinsic part of the turbulent cascade  \citep{Lazarian2020}. 

In a re-connection event, magnetic field lines break and reconnect, leading to their topological change. The resultant strong curved magnetic tension can slingshot plasma to the Alfv\'en speed. Such local kinetic energy injection could be an important driving force for high-turbulent motion, as seen in the ISM of the GC. The re-connection also helps to shape the global magnetic field topology of the ISM.  As shown in MHD simulations \citep[e.g.,][]{Hanasz2002}, the formation of large-scale poloidal magnetic field flux tubes in a turbulent medium requires the re-connection. In analog to solar flares, we may call such re-connection phenomena, driven by underlying interstellar turbulent motion, as "galactic flares" \citep{Sturrock1980}. 

In the extreme environment of the GC, magnetic field may play a key (if not dominant) role in regulating the thermal and dynamical properties, as well as the global structure of the ISM.
CRs from re-connection events can transport released magnetic field energy into surrounding medium. Such ``in-situ'' produced CRs could be responsible for the heating of the ubiqious diffuse warm molecular gas with $T \sim 200$~K,  as recently revealed by observations of H$_{3}^+$ absorption lines at 3.5-4.0\micron\  \citep{Wiener2013,Oka2019}.  The magnetic field energy may also be stored in global outflows. In this case, later release of the energy via re-connection may greatly influence the emission and the overall energetics of outflows far away from the GC.
Lines of evidence for various large-scale bipolar structures around the GC have been revealed across the spectrum:  e.g., the so-called Fermi Bubbles  in $\gamma$-ray \citep{Su2010}, the {\sl WMAP} ``haze'' in microwave, and  large-scale diffuse emissions in (polarized) radio, mid-IR and X-ray  \citep[e.g., ][]{Bland-Hawthorn2003,Carretti2013}. Kinematical information has also been obtained  from observations of UV absorption lines in the spectra of stars at various distances, as well as background AGNs, and from  HI  21~cm line emission mapping, strongly suggesting that clouds with velocities up to $\sim 10^{3} {\rm~km~s^{-1}}$ have been accelerated out from the GC, probably episodically \citep[e.g.,][]{DiTeodoro2018,Ashley2020}. Indeed, the re-connection has been proposed as a mechanism for CR acceleration to explain non-thermal emission from such large-scale structures as the Fermi Bubbles \citep[e.g.,][]{Su2010}. 

\section{Summary and Conclusions}\label{s:sum}

We have mapped a $2^\circ \times 4^\circ$ field of the Galactic center/bulge  with \chandra/ACIS-I observations, based on a dedicated gap-filling program, as well as archival data. We present maps in various bands after cleaning artifacts (mostly readout streaks and some very extended dust scattering halos of bright sources) and removing non-X-ray background contributions. 
Discrete sources detected in individual observations are merged together and are excised from the data to produce diffuse X-ray emission maps. 
We have conducted some preliminary analysis of these maps and a multi-wavelength comparison, mainly with the recent \meerkat\  survey of the GC, to provide insights into the nature of   various X-ray and radio structures, as well as their relationship. Our main results and conclusions as follows: 

\begin{itemize}

\item A pair of diffuse X-ray-emitting plumes, apparently emanating from the GC, are revealed at arcsecond resolution and with minimum confusion with discrete sources. The Galactic south plume is centrally-filled and collimated and is well enclosed by a radio lobe discovered recently by \meerkat. This association indicates that the diffuse hot plasma in the plume is confined by a strongly magnetized medium. In contrast, the Galactic north plume is spatially offset  from the well-known GCL, consistent with the proposition that this radio lobe is a foreground HII region.  An accurate localization and mapping of this HII region will be important to the determination of its role as a scatter screen of radio signals, especially those from distant pulsars  \citep{Sicheneder2017}, and to a comprehensive study of interstellar structures associated with the GC.

\item We highlight the spatial and spectral properties of an X-ray thread \xs\ that is embedded in a locally swelled radio filament, which is oriented nearly vertically with regard to the Galactic plane and is quite far off the CMZ. The X-ray spectrum of this thread can be characterized by either a power law or an optically thin thermal plasma with a temperature $\gtrsim 3$~keV. The combination of the radio and X-ray characteristics suggests that the association represents an ongoing re-connection event of antiparallel magnetic field flux tubes. \xs, together with another similar X-ray thread/NTF association \rxs,  will potentially be excellent sites to seriously learn about the interstellar re-connection astrophysics.    

\item Major re-connection events tend to occur at outer boundaries of expanding plumes or supershells of dense gas, where strong magnetic fields of different directions collide or get twisted. Detected X-ray thread/NTF associations such as \xs\  and \rxs\ probably represent only the tip of the re-connection iceberg in the GC. Most re-connection events likely produce too faint and/or too diffuse X-ray emission to be detected individually with existing observations. Re-connection events may play a major role in producing high-temperature optically-thin thermal plasma, accelerating CRs, driving turbulence and regulating global interstellar structures in and beyond the GC. 

\end{itemize}

These results and conclusions demonstrate the importance of studying high-energy phenomena and processes in the GC  and their impact on larger-scale environments over a broad range of spatial or time scales. It is also clear that the GC is a truly complex system, involving not only interplay among various stellar and interstellar components, plus Sgr A*, but also inflows and outflows,  multiple energy sources, as well as heating and cooling mechanisms. Confusion with foreground and potentially background and embedded features can also be serious. A comprehensive study of the GC with this complexity will require truly a multi-wavelength approach, together with dedicated theoretical and computer simulations. 
Ultimately, what we learn from the GC ecosystem and its connection to larger scale structures will provide us with insights into the working of similar extreme regions in other galaxies. 

\section*{Acknowledgments}

We thank the referee for useful comments and appreciate the feedback offered by Mark Morris on the work, which was supported by the National Aeronautics and Space Administration through \chandra\ Award Number GO9-20023X issued by the \chandra\ X-ray Center, which is operated by the Smithsonian Astrophysical Observatory for and on behalf of the National Aeronautics Space Administration under contract NAS8-03060, and by the ADAP grant NNX17AL67G. 
This research has made use of data obtained from the \chandra\ Data Archive and software provided by the \chandra\ X-ray Center (CXC) in the application packages CIAO. The data include all ACIS-I observations taken before 2020 and within the field of view of Fig.~1A.

\section*{DATA  AVAILABILITY}

The \chandra\ data as described in \S~\ref{s:obs} are available in the \chandra\ data archive  (https://asc.harvard.edu/cda/). Processed data products underlying this article will be shared on reasonable request to the  author.

\bibliographystyle{mnras}
\bibliography{export-bibtex_total} 

\begin{thebibliography}{}
\makeatletter
\relax
\def\mn@urlcharsother{\let\do\@makeother \do\$\do\&\do\#\do\^\do\_\do\%\do\~}
\def\mn@doi{\begingroup\mn@urlcharsother \@ifnextchar [ {\mn@doi@}
  {\mn@doi@[]}}
\def\mn@doi@[#1]#2{\def\@tempa{#1}\ifx\@tempa\@empty \href
  {http://dx.doi.org/#2} {doi:#2}\else \href {http://dx.doi.org/#2} {#1}\fi
  \endgroup}
\def\mn@eprint#1#2{\mn@eprint@#1:#2::\@nil}
\def\mn@eprint@arXiv#1{\href {http://arxiv.org/abs/#1} {{\tt arXiv:#1}}}
\def\mn@eprint@dblp#1{\href {http://dblp.uni-trier.de/rec/bibtex/#1.xml}
  {dblp:#1}}
\def\mn@eprint@#1:#2:#3:#4\@nil{\def\@tempa {#1}\def\@tempb {#2}\def\@tempc
  {#3}\ifx \@tempc \@empty \let \@tempc \@tempb \let \@tempb \@tempa \fi \ifx
  \@tempb \@empty \def\@tempb {arXiv}\fi \@ifundefined
  {mn@eprint@\@tempb}{\@tempb:\@tempc}{\expandafter \expandafter \csname
  mn@eprint@\@tempb\endcsname \expandafter{\@tempc}}}

\bibitem[\protect\citeauthoryear{{Aharonian} et~al.,}{{Aharonian}
  et~al.}{2006}]{Aharonian2006}
{Aharonian} F.,  et~al., 2006, \mn@doi [\nat] {10.1038/nature04467}, \href
  {https://ui.adsabs.harvard.edu/abs/2006Natur.439..695A} {439, 695}

\bibitem[\protect\citeauthoryear{{Alves}, {Calabretta}, {Davies}, {Dickinson},
  {Staveley-Smith}, {Davis}, {Chen}  \& {Barr}}{{Alves}
  et~al.}{2015}]{Alves2015}
{Alves} M. I.~R.,  {Calabretta} M.,  {Davies} R.~D.,  {Dickinson} C.,
  {Staveley-Smith} L.,  {Davis} R.~J.,  {Chen} T.,   {Barr} A.,  2015, \mn@doi
  [\mnras] {10.1093/mnras/stv751}, \href
  {https://ui.adsabs.harvard.edu/abs/2015MNRAS.450.2025A} {450, 2025}

\bibitem[\protect\citeauthoryear{{Anders} \& {Grevesse}}{{Anders} \&
  {Grevesse}}{1989}]{Anders1989}
{Anders} E.,  {Grevesse} N.,  1989, \mn@doi [\gca]
  {10.1016/0016-7037(89)90286-X}, \href
  {https://ui.adsabs.harvard.edu/abs/1989GeCoA..53..197A} {53, 197}

\bibitem[\protect\citeauthoryear{{Armillotta}, {Krumholz}  \& {Di
  Teodoro}}{{Armillotta} et~al.}{2020}]{Armillotta2020}
{Armillotta} L.,  {Krumholz} M.~R.,   {Di Teodoro} E.~M.,  2020, \mn@doi
  [\mnras] {10.1093/mnras/staa469}, \href
  {https://ui.adsabs.harvard.edu/abs/2020MNRAS.493.5273A} {493, 5273}

\bibitem[\protect\citeauthoryear{{Ashley}, {Fox}, {Jenkins}, {Wakker},
  {Bordoloi}, {Lockman}, {Savage}  \& {Karim}}{{Ashley}
  et~al.}{2020}]{Ashley2020}
{Ashley} T.,  {Fox} A.~J.,  {Jenkins} E.~B.,  {Wakker} B.~P.,  {Bordoloi} R.,
  {Lockman} F.~J.,  {Savage} B.~D.,   {Karim} T.,  2020, \mn@doi [\apj]
  {10.3847/1538-4357/ab9ff8}, \href
  {https://ui.adsabs.harvard.edu/abs/2020ApJ...898..128A} {898, 128}

\bibitem[\protect\citeauthoryear{{Banda-Barrag{\'a}n}, {Parkin}, {Federrath},
  {Crocker}  \& {Bicknell}}{{Banda-Barrag{\'a}n}
  et~al.}{2016}]{Banda-Barragan2016}
{Banda-Barrag{\'a}n} W.~E.,  {Parkin} E.~R.,  {Federrath} C.,  {Crocker} R.~M.,
    {Bicknell} G.~V.,  2016, \mn@doi [\mnras] {10.1093/mnras/stv2405}, \href
  {https://ui.adsabs.harvard.edu/abs/2016MNRAS.455.1309B} {455, 1309}

\bibitem[\protect\citeauthoryear{{Bandiera}}{{Bandiera}}{2008}]{Bandiera2008}
{Bandiera} R.,  2008, \mn@doi [\aap] {10.1051/0004-6361:200810666}, \href
  {https://ui.adsabs.harvard.edu/abs/2008A&A...490L...3B} {490, L3}

\bibitem[\protect\citeauthoryear{{Barkov} \& {Lyutikov}}{{Barkov} \&
  {Lyutikov}}{2019}]{Barkov2019b}
{Barkov} M.~V.,  {Lyutikov} M.,  2019, \mn@doi [\mnras]
  {10.1093/mnrasl/slz124}, \href
  {https://ui.adsabs.harvard.edu/abs/2019MNRAS.489L..28B} {489, L28}

\bibitem[\protect\citeauthoryear{{Barkov}, {Lyutikov}, {Klingler}  \&
  {Bordas}}{{Barkov} et~al.}{2019}]{Barkov2019a}
{Barkov} M.~V.,  {Lyutikov} M.,  {Klingler} N.,   {Bordas} P.,  2019, \mn@doi
  [\mnras] {10.1093/mnras/stz521}, \href
  {https://ui.adsabs.harvard.edu/abs/2019MNRAS.485.2041B} {485, 2041}

\bibitem[\protect\citeauthoryear{{Barnes}, {Longmore}, {Battersby}, {Bally},
  {Kruijssen}, {Henshaw}  \& {Walker}}{{Barnes} et~al.}{2017}]{Barnes2017}
{Barnes} A.~T.,  {Longmore} S.~N.,  {Battersby} C.,  {Bally} J.,  {Kruijssen}
  J.~M.~D.,  {Henshaw} J.~D.,   {Walker} D.~L.,  2017, \mn@doi [\mnras]
  {10.1093/mnras/stx941}, \href
  {https://ui.adsabs.harvard.edu/abs/2017MNRAS.469.2263B} {469, 2263}

\bibitem[\protect\citeauthoryear{{Benjamin}}{{Benjamin}}{2020}]{Benjamin2020}
{Benjamin} R.~A.,  2020, in preparation, private communications

\bibitem[\protect\citeauthoryear{{Bicknell} \& {Li}}{{Bicknell} \&
  {Li}}{2001}]{Bicknell2001}
{Bicknell} G.~V.,  {Li} J.,  2001, \mn@doi [\apjl] {10.1086/318928}, \href
  {https://ui.adsabs.harvard.edu/abs/2001ApJ...548L..69B} {548, L69}

\bibitem[\protect\citeauthoryear{{Bland-Hawthorn} \& {Cohen}}{{Bland-Hawthorn}
  \& {Cohen}}{2003}]{Bland-Hawthorn2003}
{Bland-Hawthorn} J.,  {Cohen} M.,  2003, \mn@doi [\apj] {10.1086/344573}, \href
  {https://ui.adsabs.harvard.edu/abs/2003ApJ...582..246B} {582, 246}

\bibitem[\protect\citeauthoryear{{Bland-Hawthorn} \&
  {Gerhard}}{{Bland-Hawthorn} \& {Gerhard}}{2016}]{Bland-Hawthorn2016}
{Bland-Hawthorn} J.,  {Gerhard} O.,  2016, \mn@doi [\araa]
  {10.1146/annurev-astro-081915-023441}, \href
  {https://ui.adsabs.harvard.edu/abs/2016ARA&A..54..529B} {54, 529}

\bibitem[\protect\citeauthoryear{{Bower} et~al.,}{{Bower}
  et~al.}{2014}]{Bower2014}
{Bower} G.~C.,  et~al., 2014, \mn@doi [\apj] {10.1088/0004-637X/790/1/1}, \href
  {https://ui.adsabs.harvard.edu/abs/2014ApJ...790....1B} {790, 1}

\bibitem[\protect\citeauthoryear{{Carretti} et~al.,}{{Carretti}
  et~al.}{2013}]{Carretti2013}
{Carretti} E.,  et~al., 2013, \mn@doi [\nat] {10.1038/nature11734}, \href
  {https://ui.adsabs.harvard.edu/abs/2013Natur.493...66C} {493, 66}

\bibitem[\protect\citeauthoryear{{Chandran}, {Cowley}  \& {Morris}}{{Chandran}
  et~al.}{2000}]{Chandran2000}
{Chandran} B. D.~G.,  {Cowley} S.~C.,   {Morris} M.,  2000, \mn@doi [\apj]
  {10.1086/308184}, \href
  {https://ui.adsabs.harvard.edu/abs/2000ApJ...528..723C} {528, 723}

\bibitem[\protect\citeauthoryear{{Di Teodoro}, {McClure-Griffiths}, {Lockman},
  {Denbo}, {Endsley}, {Ford}  \& {Harrington}}{{Di Teodoro}
  et~al.}{2018}]{DiTeodoro2018}
{Di Teodoro} E.~M.,  {McClure-Griffiths} N.~M.,  {Lockman} F.~J.,  {Denbo}
  S.~R.,  {Endsley} R.,  {Ford} H.~A.,   {Harrington} K.,  2018, \mn@doi [\apj]
  {10.3847/1538-4357/aaad6a}, \href
  {https://ui.adsabs.harvard.edu/abs/2018ApJ...855...33D} {855, 33}

\bibitem[\protect\citeauthoryear{{Figer} et~al.,}{{Figer}
  et~al.}{2002}]{Figer2002}
{Figer} D.~F.,  et~al., 2002, \mn@doi [\apj] {10.1086/344154}, \href
  {https://ui.adsabs.harvard.edu/abs/2002ApJ...581..258F} {581, 258}

\bibitem[\protect\citeauthoryear{{Florido-Llinas}, {Nieves-Chinchilla}  \&
  {Linton}}{{Florido-Llinas} et~al.}{2020}]{Florido-Llinas2020}
{Florido-Llinas} M.,  {Nieves-Chinchilla} T.,   {Linton} M.~G.,  2020, \mn@doi
  [\solphys] {10.1007/s11207-020-01687-z}, \href
  {https://ui.adsabs.harvard.edu/abs/2020SoPh..295..118F} {295, 118}

\bibitem[\protect\citeauthoryear{{Furth}, {Killeen}  \& {Rosenbluth}}{{Furth}
  et~al.}{1963}]{Furth1963}
{Furth} H.~P.,  {Killeen} J.,   {Rosenbluth} M.~N.,  1963, \mn@doi [Physics of
  Fluids] {10.1063/1.1706761}, \href
  {https://ui.adsabs.harvard.edu/abs/1963PhFl....6..459F} {6, 459}

\bibitem[\protect\citeauthoryear{{Gaensler}, {van der Swaluw}, {Camilo},
  {Kaspi}, {Baganoff}, {Yusef-Zadeh}  \& {Manchester}}{{Gaensler}
  et~al.}{2004}]{Gaensler2004}
{Gaensler} B.~M.,  {van der Swaluw} E.,  {Camilo} F.,  {Kaspi} V.~M.,
  {Baganoff} F.~K.,  {Yusef-Zadeh} F.,   {Manchester} R.~N.,  2004, \mn@doi
  [\apj] {10.1086/424906}, \href
  {https://ui.adsabs.harvard.edu/abs/2004ApJ...616..383G} {616, 383}

\bibitem[\protect\citeauthoryear{{Ginsburg} et~al.,}{{Ginsburg}
  et~al.}{2016}]{Ginsburg2016}
{Ginsburg} A.,  et~al., 2016, \mn@doi [\aap] {10.1051/0004-6361/201526100},
  \href {https://ui.adsabs.harvard.edu/abs/2016A&A...586A..50G} {586, A50}

\bibitem[\protect\citeauthoryear{{H.~E.~S.~S. Collaboration}
  et~al.,}{{H.~E.~S.~S. Collaboration}
  et~al.}{2018}]{H.E.S.S.Collaboration2018}
{H.~E.~S.~S. Collaboration} et~al., 2018, \mn@doi [\aap]
  {10.1051/0004-6361/201730824}, \href
  {https://ui.adsabs.harvard.edu/abs/2018A&A...612A...9H} {612, A9}

\bibitem[\protect\citeauthoryear{{Hanasz}, {Otmianowska-Mazur}  \&
  {Lesch}}{{Hanasz} et~al.}{2002}]{Hanasz2002}
{Hanasz} M.,  {Otmianowska-Mazur} K.,   {Lesch} H.,  2002, \mn@doi [\aap]
  {10.1051/0004-6361:20020228}, \href
  {https://ui.adsabs.harvard.edu/abs/2002A&A...386..347H} {386, 347}

\bibitem[\protect\citeauthoryear{{Heyvaerts}, {Norman}  \&
  {Pudritz}}{{Heyvaerts} et~al.}{1988}]{Heyvaerts1988}
{Heyvaerts} J.,  {Norman} C.,   {Pudritz} R.~E.,  1988, \mn@doi [\apj]
  {10.1086/166506}, \href
  {https://ui.adsabs.harvard.edu/abs/1988ApJ...330..718H} {330, 718}

\bibitem[\protect\citeauthoryear{{Heywood} et~al.,}{{Heywood}
  et~al.}{2019}]{Heywood2019}
{Heywood} I.,  et~al., 2019, \mn@doi [\nat] {10.1038/s41586-019-1532-5}, \href
  {https://ui.adsabs.harvard.edu/abs/2019Natur.573..235H} {573, 235}

\bibitem[\protect\citeauthoryear{{Hong}, {van den Berg}, {Grindlay}  \&
  {Laycock}}{{Hong} et~al.}{2009}]{Hong2009}
{Hong} J.~S.,  {van den Berg} M.,  {Grindlay} J.~E.,   {Laycock} S.,  2009,
  \mn@doi [\apj] {10.1088/0004-637X/706/1/223}, \href
  {https://ui.adsabs.harvard.edu/abs/2009ApJ...706..223H} {706, 223}

\bibitem[\protect\citeauthoryear{{Ieda}, {Machida}, {Mukai}, {Saito},
  {Yamamoto}, {Nishida}, {Terasawa}  \& {Kokubun}}{{Ieda}
  et~al.}{1998}]{Ieda1998}
{Ieda} A.,  {Machida} S.,  {Mukai} T.,  {Saito} Y.,  {Yamamoto} T.,  {Nishida}
  A.,  {Terasawa} T.,   {Kokubun} S.,  1998, \mn@doi [\jgr]
  {10.1029/97JA03240}, \href
  {https://ui.adsabs.harvard.edu/abs/1998JGR...103.4453I} {103, 4453}

\bibitem[\protect\citeauthoryear{{Johnson}, {Dong}  \& {Wang}}{{Johnson}
  et~al.}{2009}]{Johnson2009}
{Johnson} S.~P.,  {Dong} H.,   {Wang} Q.~D.,  2009, \mn@doi [\mnras]
  {10.1111/j.1365-2966.2009.15362.x}, \href
  {https://ui.adsabs.harvard.edu/abs/2009MNRAS.399.1429J} {399, 1429}

\bibitem[\protect\citeauthoryear{{Jonker} et~al.,}{{Jonker}
  et~al.}{2014}]{Jonker2014}
{Jonker} P.~G.,  et~al., 2014, \mn@doi [\apjs] {10.1088/0067-0049/210/2/18},
  \href {https://ui.adsabs.harvard.edu/abs/2014ApJS..210...18J} {210, 18}

\bibitem[\protect\citeauthoryear{{Koyama}}{{Koyama}}{2018}]{Koyama2018}
{Koyama} K.,  2018, \mn@doi [\pasj] {10.1093/pasj/psx084}, \href
  {https://ui.adsabs.harvard.edu/abs/2018PASJ...70R...1K} {70, R1}

\bibitem[\protect\citeauthoryear{{Kruijssen} \& {Longmore}}{{Kruijssen} \&
  {Longmore}}{2013}]{Kruijssen2013}
{Kruijssen} J.~M.~D.,  {Longmore} S.~N.,  2013, \mn@doi [\mnras]
  {10.1093/mnras/stt1634}, \href
  {https://ui.adsabs.harvard.edu/abs/2013MNRAS.435.2598K} {435, 2598}

\bibitem[\protect\citeauthoryear{{Kruijssen}, {Dale}  \&
  {Longmore}}{{Kruijssen} et~al.}{2015}]{Kruijssen2015}
{Kruijssen} J.~M.~D.,  {Dale} J.~E.,   {Longmore} S.~N.,  2015, \mn@doi
  [\mnras] {10.1093/mnras/stu2526}, \href
  {https://ui.adsabs.harvard.edu/abs/2015MNRAS.447.1059K} {447, 1059}

\bibitem[\protect\citeauthoryear{{LaRosa}, {Lazio}  \& {Kassim}}{{LaRosa}
  et~al.}{2001}]{larosa2001}
{LaRosa} T.~N.,  {Lazio} T. J.~W.,   {Kassim} N.~E.,  2001, \mn@doi [\apj]
  {10.1086/323793}, \href
  {https://ui.adsabs.harvard.edu/abs/2001ApJ...563..163L} {563, 163}

\bibitem[\protect\citeauthoryear{{LaRosa}, {Nord}, {Lazio}  \&
  {Kassim}}{{LaRosa} et~al.}{2004}]{LaRosa2004}
{LaRosa} T.~N.,  {Nord} M.~E.,  {Lazio} T. J.~W.,   {Kassim} N.~E.,  2004,
  \mn@doi [\apj] {10.1086/383233}, \href
  {https://ui.adsabs.harvard.edu/abs/2004ApJ...607..302L} {607, 302}

\bibitem[\protect\citeauthoryear{{Law}, {Backer}, {Yusef-Zadeh}  \&
  {Maddalena}}{{Law} et~al.}{2009}]{Law2009}
{Law} C.~J.,  {Backer} D.,  {Yusef-Zadeh} F.,   {Maddalena} R.,  2009, \mn@doi
  [\apj] {10.1088/0004-637X/695/2/1070}, \href
  {https://ui.adsabs.harvard.edu/abs/2009ApJ...695.1070L} {695, 1070}

\bibitem[\protect\citeauthoryear{{Lazarian} \& {Desiati}}{{Lazarian} \&
  {Desiati}}{2010}]{Lazarian2010}
{Lazarian} A.,  {Desiati} P.,  2010, \mn@doi [\apj]
  {10.1088/0004-637X/722/1/188}, \href
  {https://ui.adsabs.harvard.edu/abs/2010ApJ...722..188L} {722, 188}

\bibitem[\protect\citeauthoryear{{Lazarian} \& {Vishniac}}{{Lazarian} \&
  {Vishniac}}{1999}]{Lazarian1999}
{Lazarian} A.,  {Vishniac} E.~T.,  1999, \mn@doi [\apj] {10.1086/307233}, \href
  {https://ui.adsabs.harvard.edu/abs/1999ApJ...517..700L} {517, 700}

\bibitem[\protect\citeauthoryear{{Lazarian}, {Eyink}, {Jafari}, {Kowal}, {Li},
  {Xu}  \& {Vishniac}}{{Lazarian} et~al.}{2020}]{Lazarian2020}
{Lazarian} A.,  {Eyink} G.~L.,  {Jafari} A.,  {Kowal} G.,  {Li} H.,  {Xu} S.,
  {Vishniac} E.~T.,  2020, \mn@doi [Physics of Plasmas] {10.1063/1.5110603},
  \href {https://ui.adsabs.harvard.edu/abs/2020PhPl...27a2305L} {27, 012305}

\bibitem[\protect\citeauthoryear{{Libralato}, {Fardal}, {Lennon}, {van der
  Marel}  \& {Bellini}}{{Libralato} et~al.}{2020}]{Libralato2020}
{Libralato} M.,  {Fardal} M.,  {Lennon} D.,  {van der Marel} R.~P.,   {Bellini}
  A.,  2020, \mn@doi [\mnras] {10.1093/mnras/staa2327}, \href
  {https://ui.adsabs.harvard.edu/abs/2020MNRAS.497.4733L} {497, 4733}

\bibitem[\protect\citeauthoryear{{Lu}, {Wang}  \& {Lang}}{{Lu}
  et~al.}{2003}]{Lu2003}
{Lu} F.~J.,  {Wang} Q.~D.,   {Lang} C.~C.,  2003, \mn@doi [\aj]
  {10.1086/375754}, \href
  {https://ui.adsabs.harvard.edu/abs/2003AJ....126..319L} {126, 319}

\bibitem[\protect\citeauthoryear{{Lu}, {Angelopoulos}, {Artemyev}, {Pritchett},
  {Sun}  \& {Slavin}}{{Lu} et~al.}{2020}]{Lu2020}
{Lu} S.,  {Angelopoulos} V.,  {Artemyev} A.~V.,  {Pritchett} P.~L.,  {Sun}
  W.~J.,   {Slavin} J.~A.,  2020, \mn@doi [\apj] {10.3847/1538-4357/abaa44},
  \href {https://ui.adsabs.harvard.edu/abs/2020ApJ...900..145L} {900, 145}

\bibitem[\protect\citeauthoryear{{Mezger}, {Duschl}  \& {Zylka}}{{Mezger}
  et~al.}{1996}]{Mezger1996}
{Mezger} P.~G.,  {Duschl} W.~J.,   {Zylka} R.,  1996, \mn@doi [\aapr]
  {10.1007/s001590050007}, \href
  {https://ui.adsabs.harvard.edu/abs/1996A&ARv...7..289M} {7, 289}

\bibitem[\protect\citeauthoryear{{Morris}}{{Morris}}{2006}]{Morris2006a}
{Morris} M.,  2006, in Journal of Physics Conference Series. pp~1--9,
  \mn@doi{10.1088/1742-6596/54/1/001}

\bibitem[\protect\citeauthoryear{{Morris}}{{Morris}}{2015}]{Morris2015}
{Morris} M.~R.,  2015, {Manifestations of the Galactic Center Magnetic Field}.
p.~391, \mn@doi{10.1007/978-3-319-10614-4_32}

\bibitem[\protect\citeauthoryear{{Morris} \& {Serabyn}}{{Morris} \&
  {Serabyn}}{1996}]{Morris1996}
{Morris} M.,  {Serabyn} E.,  1996, \mn@doi [\araa]
  {10.1146/annurev.astro.34.1.645}, \href
  {https://ui.adsabs.harvard.edu/abs/1996ARA&A..34..645M} {34, 645}

\bibitem[\protect\citeauthoryear{{Morris}, {Uchida}  \& {Do}}{{Morris}
  et~al.}{2006}]{Morris2006b}
{Morris} M.,  {Uchida} K.,   {Do} T.,  2006, \mn@doi [\nat]
  {10.1038/nature04554}, \href
  {https://ui.adsabs.harvard.edu/abs/2006Natur.440..308M} {440, 308}

\bibitem[\protect\citeauthoryear{{Muno} et~al.,}{{Muno}
  et~al.}{2009}]{Muno2009}
{Muno} M.~P.,  et~al., 2009, \mn@doi [\apjs] {10.1088/0067-0049/181/1/110},
  \href {https://ui.adsabs.harvard.edu/abs/2009ApJS..181..110M} {181, 110}

\bibitem[\protect\citeauthoryear{{Nagoshi} et~al.,}{{Nagoshi}
  et~al.}{2019}]{Nagoshi2019}
{Nagoshi} H.,  et~al., 2019, \mn@doi [\pasj] {10.1093/pasj/psz060}, \href
  {https://ui.adsabs.harvard.edu/abs/2019PASJ...71...80N} {71, 80}

\bibitem[\protect\citeauthoryear{{Nakashima}, {Nobukawa}, {Uchida}, {Tanaka},
  {Tsuru}, {Koyama}, {Murakami}  \& {Uchiyama}}{{Nakashima}
  et~al.}{2013}]{Nakashima2013}
{Nakashima} S.,  {Nobukawa} M.,  {Uchida} H.,  {Tanaka} T.,  {Tsuru} T.~G.,
  {Koyama} K.,  {Murakami} H.,   {Uchiyama} H.,  2013, \mn@doi [\apj]
  {10.1088/0004-637X/773/1/20}, \href
  {https://ui.adsabs.harvard.edu/abs/2013ApJ...773...20N} {773, 20}

\bibitem[\protect\citeauthoryear{{Nakashima}, {Koyama}, {Wang}  \&
  {Enokiya}}{{Nakashima} et~al.}{2019}]{Nakashima2019}
{Nakashima} S.,  {Koyama} K.,  {Wang} Q.~D.,   {Enokiya} R.,  2019, \mn@doi
  [\apj] {10.3847/1538-4357/ab0d82}, \href
  {https://ui.adsabs.harvard.edu/abs/2019ApJ...875...32N} {875, 32}

\bibitem[\protect\citeauthoryear{{Nishiyama} et~al.,}{{Nishiyama}
  et~al.}{2013}]{Nishiyama2013}
{Nishiyama} S.,  et~al., 2013, \mn@doi [\apjl] {10.1088/2041-8205/769/2/L28},
  \href {https://ui.adsabs.harvard.edu/abs/2013ApJ...769L..28N} {769, L28}

\bibitem[\protect\citeauthoryear{{Nogueras-Lara} et~al.,}{{Nogueras-Lara}
  et~al.}{2018}]{Nogueras-Lara2018}
{Nogueras-Lara} F.,  et~al., 2018, \mn@doi [\aap]
  {10.1051/0004-6361/201833518}, \href
  {https://ui.adsabs.harvard.edu/abs/2018A&A...620A..83N} {620, A83}

\bibitem[\protect\citeauthoryear{{Oka}, {Geballe}, {Goto}, {Usuda}, {Benjamin},
  {McCall}  \& {Indriolo}}{{Oka} et~al.}{2019}]{Oka2019}
{Oka} T.,  {Geballe} T.~R.,  {Goto} M.,  {Usuda} T.,  {Benjamin} {McCall} J.,
  {Indriolo} N.,  2019, \mn@doi [\apj] {10.3847/1538-4357/ab3647}, \href
  {https://ui.adsabs.harvard.edu/abs/2019ApJ...883...54O} {883, 54}

\bibitem[\protect\citeauthoryear{{Park} et~al.,}{{Park}
  et~al.}{2005}]{Park2005}
{Park} S.,  et~al., 2005, \mn@doi [\apj] {10.1086/432639}, \href
  {https://ui.adsabs.harvard.edu/abs/2005ApJ...631..964P} {631, 964}

\bibitem[\protect\citeauthoryear{{Ponti} et~al.,}{{Ponti}
  et~al.}{2015}]{Ponti2015}
{Ponti} G.,  et~al., 2015, \mn@doi [\mnras] {10.1093/mnras/stv1331}, \href
  {https://ui.adsabs.harvard.edu/abs/2015MNRAS.453..172P} {453, 172}

\bibitem[\protect\citeauthoryear{{Ponti} et~al.,}{{Ponti}
  et~al.}{2019}]{Ponti2019}
{Ponti} G.,  et~al., 2019, \mn@doi [\nat] {10.1038/s41586-019-1009-6}, \href
  {https://ui.adsabs.harvard.edu/abs/2019Natur.567..347P} {567, 347}

\bibitem[\protect\citeauthoryear{{Possenti}, {Cerutti}, {Colpi}  \&
  {Mereghetti}}{{Possenti} et~al.}{2002}]{Possenti2002}
{Possenti} A.,  {Cerutti} R.,  {Colpi} M.,   {Mereghetti} S.,  2002, \mn@doi
  [\aap] {10.1051/0004-6361:20020472}, \href
  {https://ui.adsabs.harvard.edu/abs/2002A&A...387..993P} {387, 993}

\bibitem[\protect\citeauthoryear{{Raymond}}{{Raymond}}{1992}]{Raymond1992}
{Raymond} J.~C.,  1992, \mn@doi [\apj] {10.1086/170892}, \href
  {https://ui.adsabs.harvard.edu/abs/1992ApJ...384..502R} {384, 502}

\bibitem[\protect\citeauthoryear{{Revnivtsev}, {Sazonov}, {Churazov}, {Forman},
  {Vikhlinin}  \& {Sunyaev}}{{Revnivtsev} et~al.}{2009}]{Revnivtsev2009}
{Revnivtsev} M.,  {Sazonov} S.,  {Churazov} E.,  {Forman} W.,  {Vikhlinin} A.,
   {Sunyaev} R.,  2009, \mn@doi [\nat] {10.1038/nature07946}, \href
  {https://ui.adsabs.harvard.edu/abs/2009Natur.458.1142R} {458, 1142}

\bibitem[\protect\citeauthoryear{{Serabyn} \& {Morris}}{{Serabyn} \&
  {Morris}}{1994}]{Serabyn1994}
{Serabyn} E.,  {Morris} M.,  1994, \mn@doi [\apjl] {10.1086/187282}, \href
  {https://ui.adsabs.harvard.edu/abs/1994ApJ...424L..91S} {424, L91}

\bibitem[\protect\citeauthoryear{{Shore} \& {LaRosa}}{{Shore} \&
  {LaRosa}}{1999}]{Shore1999}
{Shore} S.~N.,  {LaRosa} T.~N.,  1999, \mn@doi [\apj] {10.1086/307601}, \href
  {https://ui.adsabs.harvard.edu/abs/1999ApJ...521..587S} {521, 587}

\bibitem[\protect\citeauthoryear{{Sicheneder} \& {Dexter}}{{Sicheneder} \&
  {Dexter}}{2017}]{Sicheneder2017}
{Sicheneder} E.,  {Dexter} J.,  2017, \mn@doi [\mnras] {10.1093/mnras/stx103},
  \href {https://ui.adsabs.harvard.edu/abs/2017MNRAS.467.3642S} {467, 3642}

\bibitem[\protect\citeauthoryear{{Sofue}}{{Sofue}}{2013}]{Sofue2013}
{Sofue} Y.,  2013, \mn@doi [\pasj] {10.1093/pasj/65.6.118}, \href
  {https://ui.adsabs.harvard.edu/abs/2013PASJ...65..118S} {65, 118}

\bibitem[\protect\citeauthoryear{{Sofue}}{{Sofue}}{2020}]{Sofue2020}
{Sofue} Y.,  2020, \mn@doi [\pasj] {10.1093/pasj/psaa011}, \href
  {https://ui.adsabs.harvard.edu/abs/2020PASJ...72L...4S} {72, L4}

\bibitem[\protect\citeauthoryear{{Sofue} \& {Fujimoto}}{{Sofue} \&
  {Fujimoto}}{1987}]{Sofue1987}
{Sofue} Y.,  {Fujimoto} M.,  1987, \mn@doi [\apjl] {10.1086/184957}, \href
  {https://ui.adsabs.harvard.edu/abs/1987ApJ...319L..73S} {319, L73}

\bibitem[\protect\citeauthoryear{{Sofue}, {Kigure}  \& {Shibata}}{{Sofue}
  et~al.}{2005}]{Sofue2005}
{Sofue} Y.,  {Kigure} H.,   {Shibata} K.,  2005, \mn@doi [\pasj]
  {10.1093/pasj/57.5.L39}, \href
  {https://ui.adsabs.harvard.edu/abs/2005PASJ...57L..39S} {57, L39}

\bibitem[\protect\citeauthoryear{{Sturrock} \& {Stern}}{{Sturrock} \&
  {Stern}}{1980}]{Sturrock1980}
{Sturrock} P.~A.,  {Stern} R.,  1980, \mn@doi [\apj] {10.1086/157962}, \href
  {https://ui.adsabs.harvard.edu/abs/1980ApJ...238...98S} {238, 98}

\bibitem[\protect\citeauthoryear{{Su}, {Slatyer}  \& {Finkbeiner}}{{Su}
  et~al.}{2010}]{Su2010}
{Su} M.,  {Slatyer} T.~R.,   {Finkbeiner} D.~P.,  2010, \mn@doi [\apj]
  {10.1088/0004-637X/724/2/1044}, \href
  {https://ui.adsabs.harvard.edu/abs/2010ApJ...724.1044S} {724, 1044}

\bibitem[\protect\citeauthoryear{{Tanuma}, {Yokoyama}, {Kudoh}, {Matsumoto},
  {Shibata}  \& {Makishima}}{{Tanuma} et~al.}{1999}]{Tanuma1999}
{Tanuma} S.,  {Yokoyama} T.,  {Kudoh} T.,  {Matsumoto} R.,  {Shibata} K.,
  {Makishima} K.,  1999, \mn@doi [\pasj] {10.1093/pasj/51.1.161}, \href
  {https://ui.adsabs.harvard.edu/abs/1999PASJ...51..161T} {51, 161}

\bibitem[\protect\citeauthoryear{{Tanuma}, {Yokoyama}, {Kudoh}  \&
  {Shibata}}{{Tanuma} et~al.}{2003}]{Tanuma2003}
{Tanuma} S.,  {Yokoyama} T.,  {Kudoh} T.,   {Shibata} K.,  2003, \mn@doi [\apj]
  {10.1086/344523}, \href
  {https://ui.adsabs.harvard.edu/abs/2003ApJ...582..215T} {582, 215}

\bibitem[\protect\citeauthoryear{{Thomas}, {Pfrommer}  \&
  {En{\ss}lin}}{{Thomas} et~al.}{2020}]{Thomas2020}
{Thomas} T.,  {Pfrommer} C.,   {En{\ss}lin} T.,  2020, \mn@doi [\apjl]
  {10.3847/2041-8213/ab7237}, \href
  {https://ui.adsabs.harvard.edu/abs/2020ApJ...890L..18T} {890, L18}

\bibitem[\protect\citeauthoryear{{Titov} \& {D{\'e}moulin}}{{Titov} \&
  {D{\'e}moulin}}{1999}]{Titov1999}
{Titov} V.~S.,  {D{\'e}moulin} P.,  1999, \aap, \href
  {https://ui.adsabs.harvard.edu/abs/1999A&A...351..707T} {351, 707}

\bibitem[\protect\citeauthoryear{{Toriumi} \& {Wang}}{{Toriumi} \&
  {Wang}}{2019}]{Toriumi2019}
{Toriumi} S.,  {Wang} H.,  2019, \mn@doi [Living Reviews in Solar Physics]
  {10.1007/s41116-019-0019-7}, \href
  {https://ui.adsabs.harvard.edu/abs/2019LRSP...16....3T} {16, 3}

\bibitem[\protect\citeauthoryear{{Tsap}, {Fedun}, {Cheremnykh}, {Stepanov},
  {Kryshtal}  \& {Kopylova}}{{Tsap} et~al.}{2020}]{Tsap2020}
{Tsap} Y.,  {Fedun} V.,  {Cheremnykh} O.,  {Stepanov} A.,  {Kryshtal} A.,
  {Kopylova} Y.,  2020, \mn@doi [\apj] {10.3847/1538-4357/abaf01}, \href
  {https://ui.adsabs.harvard.edu/abs/2020ApJ...901...99T} {901, 99}

\bibitem[\protect\citeauthoryear{{Tsuboi}, {Tsutsumi}, {Kitamura}, {Miyawaki},
  {Miyazaki}  \& {Miyoshi}}{{Tsuboi} et~al.}{2020}]{tsuboi2020}
{Tsuboi} M.,  {Tsutsumi} T.,  {Kitamura} Y.,  {Miyawaki} R.,  {Miyazaki} A.,
  {Miyoshi} M.,  2020, \mn@doi [\pasj] {10.1093/pasj/psaa077}, \href
  {https://ui.adsabs.harvard.edu/abs/2020PASJ..tmp..231T} {}

\bibitem[\protect\citeauthoryear{{Uchida}, {Morris}, {Serabyn}  \&
  {Bally}}{{Uchida} et~al.}{1994}]{Uchida1994}
{Uchida} K.~I.,  {Morris} M.~R.,  {Serabyn} E.,   {Bally} J.,  1994, \mn@doi
  [\apj] {10.1086/173667}, \href
  {https://ui.adsabs.harvard.edu/abs/1994ApJ...421..505U} {421, 505}

\bibitem[\protect\citeauthoryear{{Wang}}{{Wang}}{2004}]{Wang2004}
{Wang} Q.~D.,  2004, \mn@doi [\apj] {10.1086/422553}, \href
  {https://ui.adsabs.harvard.edu/abs/2004ApJ...612..159W} {612, 159}

\bibitem[\protect\citeauthoryear{{Wang}, {Gotthelf}  \& {Lang}}{{Wang}
  et~al.}{2002a}]{Wang2002}
{Wang} Q.~D.,  {Gotthelf} E.~V.,   {Lang} C.~C.,  2002a, \mn@doi [\nat]
  {10.1038/415148a}, \href
  {https://ui.adsabs.harvard.edu/abs/2002Natur.415..148W} {415, 148}

\bibitem[\protect\citeauthoryear{{Wang}, {Lu}  \& {Lang}}{{Wang}
  et~al.}{2002b}]{Wang2002a}
{Wang} Q.~D.,  {Lu} F.,   {Lang} C.~C.,  2002b, \mn@doi [\apj]
  {10.1086/344401}, \href
  {https://ui.adsabs.harvard.edu/abs/2002ApJ...581.1148W} {581, 1148}

\bibitem[\protect\citeauthoryear{{Wang}, {Lu}  \& {Gotthelf}}{{Wang}
  et~al.}{2006}]{Wang2006}
{Wang} Q.~D.,  {Lu} F.~J.,   {Gotthelf} E.~V.,  2006, \mn@doi [\mnras]
  {10.1111/j.1365-2966.2006.09998.x}, \href
  {https://ui.adsabs.harvard.edu/abs/2006MNRAS.367..937W} {367, 937}

\bibitem[\protect\citeauthoryear{{Wang} et~al.,}{{Wang}
  et~al.}{2013}]{Wang2013}
{Wang} Q.~D.,  et~al., 2013, \mn@doi [Science] {10.1126/science.1240755}, \href
  {https://ui.adsabs.harvard.edu/abs/2013Sci...341..981W} {341, 981}

\bibitem[\protect\citeauthoryear{{We{\.z}gowiec}, {Ehle}, {Soida}, {Dettmar},
  {Beck}  \& {Urbanik}}{{We{\.z}gowiec} et~al.}{2020}]{Wezgowiec2020}
{We{\.z}gowiec} M.,  {Ehle} M.,  {Soida} M.,  {Dettmar} R.~J.,  {Beck} R.,
  {Urbanik} M.,  2020, \mn@doi [\aap] {10.1051/0004-6361/202037842}, \href
  {https://ui.adsabs.harvard.edu/abs/2020A&A...640A.109W} {640, A109}

\bibitem[\protect\citeauthoryear{{Wiener}, {Zweibel}  \& {Oh}}{{Wiener}
  et~al.}{2013}]{Wiener2013}
{Wiener} J.,  {Zweibel} E.~G.,   {Oh} S.~P.,  2013, \mn@doi [\apj]
  {10.1088/0004-637X/767/1/87}, \href
  {https://ui.adsabs.harvard.edu/abs/2013ApJ...767...87W} {767, 87}

\bibitem[\protect\citeauthoryear{{Yamauchi}, {Shimizu}, {Nakashima},
  {Nobukawa}, {Tsuru}  \& {Koyama}}{{Yamauchi} et~al.}{2014}]{Yamauchi2014}
{Yamauchi} S.,  {Shimizu} M.,  {Nakashima} S.,  {Nobukawa} M.,  {Tsuru} T.~G.,
   {Koyama} K.,  2014, \mn@doi [\pasj] {10.1093/pasj/psu122}, \href
  {https://ui.adsabs.harvard.edu/abs/2014PASJ...66..125Y} {66, 125}

\bibitem[\protect\citeauthoryear{{Yamauchi}, {Shimizu}, {Nobukawa}, {Nobukawa},
  {Uchiyama}  \& {Koyama}}{{Yamauchi} et~al.}{2018}]{Yamauchi2018}
{Yamauchi} S.,  {Shimizu} M.,  {Nobukawa} M.,  {Nobukawa} K.~K.,  {Uchiyama}
  H.,   {Koyama} K.,  2018, \mn@doi [\pasj] {10.1093/pasj/psy077}, \href
  {https://ui.adsabs.harvard.edu/abs/2018PASJ...70...82Y} {70, 82}

\bibitem[\protect\citeauthoryear{{Yusef-Zadeh} \& {Morris}}{{Yusef-Zadeh} \&
  {Morris}}{1987a}]{Yusef-Zadeh1987b}
{Yusef-Zadeh} F.,  {Morris} M.,  1987a, \mn@doi [\aj] {10.1086/114555}, \href
  {https://ui.adsabs.harvard.edu/abs/1987AJ.....94.1178Y} {94, 1178}

\bibitem[\protect\citeauthoryear{{Yusef-Zadeh} \& {Morris}}{{Yusef-Zadeh} \&
  {Morris}}{1987b}]{Yusef-Zadeh1987a}
{Yusef-Zadeh} F.,  {Morris} M.,  1987b, \mn@doi [\apj] {10.1086/165767}, \href
  {https://ui.adsabs.harvard.edu/abs/1987ApJ...322..721Y} {322, 721}

\bibitem[\protect\citeauthoryear{{Yusef-Zadeh} \& {Wardle}}{{Yusef-Zadeh} \&
  {Wardle}}{2019}]{Yusef-Zadeh2019}
{Yusef-Zadeh} F.,  {Wardle} M.,  2019, \mn@doi [\mnras]
  {10.1093/mnrasl/slz134}, \href
  {https://ui.adsabs.harvard.edu/abs/2019MNRAS.490L...1Y} {490, L1}

\bibitem[\protect\citeauthoryear{{Yusef-Zadeh}, {Morris}  \&
  {Chance}}{{Yusef-Zadeh} et~al.}{1984}]{Yusef-Zadeh1984}
{Yusef-Zadeh} F.,  {Morris} M.,   {Chance} D.,  1984, \mn@doi [\nat]
  {10.1038/310557a0}, \href
  {https://ui.adsabs.harvard.edu/abs/1984Natur.310..557Y} {310, 557}

\bibitem[\protect\citeauthoryear{{Yusef-Zadeh}, {Wardle}  \&
  {Parastaran}}{{Yusef-Zadeh} et~al.}{1997}]{Yusef-Zadeh1997}
{Yusef-Zadeh} F.,  {Wardle} M.,   {Parastaran} P.,  1997, \mn@doi [\apjl]
  {10.1086/310484}, \href
  {https://ui.adsabs.harvard.edu/abs/1997ApJ...475L.119Y} {475, L119}

\bibitem[\protect\citeauthoryear{{Zhang} et~al.,}{{Zhang}
  et~al.}{2020}]{Zhang2020}
{Zhang} S.,  et~al., 2020, \mn@doi [\apj] {10.3847/1538-4357/ab7dc1}, \href
  {https://ui.adsabs.harvard.edu/abs/2020ApJ...893....3Z} {893, 3}

\bibitem[\protect\citeauthoryear{{Zweibel} \& {Yamada}}{{Zweibel} \&
  {Yamada}}{2009}]{Zweibel2009}
{Zweibel} E.~G.,  {Yamada} M.,  2009, \mn@doi [\araa]
  {10.1146/annurev-astro-082708-101726}, \href
  {https://ui.adsabs.harvard.edu/abs/2009ARA&A..47..291Z} {47, 291}

\makeatother
\end{thebibliography}
\end{document}